\documentclass[12pt,a4paper,final]{iopart}

\usepackage{iopams}

\usepackage{graphicx}
\usepackage{dcolumn}
\usepackage[usenames,dvipsnames,svgnames,table]{xcolor}
\usepackage[colorlinks=true,
            linkcolor=blue,
            urlcolor=black,
            citecolor=blue]{hyperref}
\usepackage{bm}
\usepackage{enumerate}
\usepackage{lipsum}
\usepackage{epstopdf}
\usepackage{float}
\usepackage[caption = false]{subfig}
\usepackage{tabularx}
\usepackage{color}

\begin{document}

\title{Tuning of non-paraxial effects of the Laguerre-Gaussian beam interacting with the two-component  Bose-Einstein condensates}

\author[cor1]{Anal Bhowmik and Sonjoy Majumder}
\address{Department of Physics, Indian Institute of Technology Kharagpur, Kharagpur-721302, India.}

\eads{\mailto{analbhowmik@phy.iitkgp.ac.in}}
\eads{\mailto{sonjoym@phy.iitkgp.ac.in}}

\begin{abstract}
We present the theory of microscopic interaction of the spin-orbit coupled focused Laguerre-Gaussian (LG) beam with the two-component Bose-Einstein condensate (BEC), composed of two hyperfine states of $^{87}$Rb in a harmonic trap. We have shown that Raman Rabi frequency distributions over the inter-component coupling identify phase separation coupling strength. A significant enhancement of side-band transitions due to non-paraxial nature of vortex beam is observed for particular values of inter-component coupling around 1.25 and 0.64 in unit of 5.5nm for  $10^5$ and $10^6$ number of  atoms, respectively. The uncertainty in the estimation of these coupling strengths is improved with the focusing angles of the beam. We discuss an experimental scheme to verify this non-paraxial effect on ultra-cold atoms.

\end{abstract}


\section{INTRODUCTION}

After the JILA experiments on two hyperfine states of $^{87}$Rb \cite{Hall1998}, the coupled Bose-Einstein condensates (BEC) became a highly useful artificial model  for  studying a wide variety of real condensed matter systems. The study of miscibility-immiscibility phase transition \cite{Jain2011} of the two-component BEC and its tunable interaction through magnetic or optical Feshbach-resonances \cite{Chin2010} provides rich insight into many-body quantum physics of the system and the origin of such phenomena.  Examples of  these quantum phenomena  are  the Kibble-Zurek mechanism \cite{Nicklas2015}, the production of dipolar molecules \cite{Molony2014}, vortex-antivortex molecules \cite{Geurts2008},  phase separation \cite{ McCarron2011, Wacker2015, Wang2016, Papp2008}, pattern formation \cite{Sabbatini2011, Hoefer2011, Hamner2011, De2014}, symmetry breaking
transitions \cite{Lee2009}, skyrmions \cite{Kawakami2012, Orlova2016}, exotic vortex lattices \cite{Kuopanportti2012}, solitary multiquantum vortices \cite{Kuopanportti2015} collective modes \cite{Barbut2014}, nonlinear dynamical excitations \cite{Mertes2007, Eto2016}, quantum turbulence \cite{Takeuchi2010}, vortex bright solitons \cite{Law2010} and vortex dynamics  in coherently coupled BEC \cite{Calderaro2017}. Diverse investigations in the above mentioned arenas of  research have been carried out on two-component BEC  using two different alkali-metal atoms \cite{Lercher2011, Pasquiou2013,  Roy2015, Lee2016, Bandyopadhyay2017} or different isotopes of same atom \cite{Sugawa2011, Inouye1998,  Tojo2010} or   same isotopes with different hyperfine states  \cite{Stenger1998, Sadler2006}.

Our recent work \cite{Anal2018} on the analysis of the structure  of  a two-component BEC with paraxial Laguerre-Gaussian (LG) beam has motivated us to study further on matter-vortices in the  BEC mixture due to non-paraxial LG beam. Although the effect of orbital angular momentum (OAM) of LG beam on the  center-of-mass (CM) motion  of atoms at BEC was experimentally demonstrated \cite{Andersen2006,Wright2008} more than a decade ago, it is the theoretical derivation of Mondal \textit{et al.} \cite{Mondal2014, Mondal2015} that provides a detailed picture of the transfer mechanism of both the orbital  and spin angular momentum (SAM) from paraxial LG beam to the internal and external motions of atoms below their recoil limit.

 However, the transfer mechanism of angular momenta from the non-paraxial LG beam to the ultra-cold atoms  is quite different compared to that of the paraxial LG beam. Unlike the latter case,  the OAM and the SAM are no longer conserved separately for the former case, in interaction with an ultra-cold atom or  molecule, but the total angular momentum (OAM+SAM) is conserved  \cite{Marrucci2006,Zhao2007}.  Our recent study \cite{Bhowmik2016}, shows that the OAM of focused LG beam can be  transferred to the electronic motion of an ultracold atom  even at the dipole transition  level.  That paper demonstrates the generation of  three possible transition channels of light-matter interaction distributing  the total angular momentum of the focused light to the internal electronic and external CM motions of atoms \cite{Bhowmik2016}.  This extra degree of freedom provides  control on the interaction as well as on the choice of the channels. Among those three channels,  two channels are comparatively weak, lets call them 'side-band' transitions. These side-band transitions channels also correspond to the transfer of the field polarization to the external motion of the atoms. In spite of their weakness, we will show the importance and  enhancement of strength of these channels in particular physical conditions.   We further study the effects of  the focusing angle of the LG beam interacting with the two-component BEC using the proper choice of the inter- and intra- component interaction strengths.  The nonparaxial vortex beams have important applications
 in different fields of science, such as trapping of atoms \cite{Bhowmik2018, Chu1986} or microparticles \cite{Ashkin1986}, optical transitions in semiconductors \cite{Quinteiro2010}, quantum information processing \cite{Beugnon2007} and cell biology \cite{Mehta1999}, etc.

The main aims of this paper are to study the effect of the non-paraxial nature of the LG beam on the two-component BEC  and its application to analyze the  structure of the density of the BEC depending on the inter-component coupling strength. Two hyperfine states of $^{87}$Rb are considered as two-component BEC here. To realize the effect quantitatively, we study  the variation of the Rabi frequencies of the two-photon stimulated Raman transitions for different focusing angles of the LG beam, which  interacts with the diverse ground state structures of the two component BEC  produced due  to the different inter- and intra-component scattering lengths.   We find that the effect of non-paraxial  LG beam is significant on the  two component BEC  for certain values of inter-component interaction at a fixed intra-component interaction strengths. 

\section{THEORY}

 In the mean field approximation, the stationary ground-state of  a dilute mixture of two-component BEC trapped in a harmonic potential at $T=0$ K is governed by coupled Gross-Pitaeveskii (GP) equations \cite{Ho1996, Jezek2001, Pu1997} 
\begin{center}
\begin{equation}
\hspace{-2.3cm}\left[-\frac{\hbar^2\nabla^2}{2m_i}+V_i(\textbf{R})+ \sum_{j=1}^{2}U_{ij}|\Psi_j(\textbf{R})|^2\right]\Psi_i(\textbf{R})=\mu_i \Psi_i(\textbf{R}),
\end{equation}
\end{center}
where $i=1$ and $2$ are indexes of components of BEC with the  normalization condition $\int |\Psi_i(\textbf{R})|^2 d\textbf{R}=N_i$. Here $N_i$, $m_i$ and $\mu_i$ denote the number of atoms, mass of the atom, and the chemical potential of the $i$-th  component of BEC.  $\Psi_i$ is the CM wavefunctions of the corresponding  component of BEC. The asymmetrical harmonic potential  is $V_i(\textbf{R})=\frac{1}{2}m_i(\omega^2_{\bot}R^2+\omega^2_{Z}Z^2)$, where  $\omega_{\bot}$ and $\omega_Z$ are trapping frequencies in the  $X-Y$ plane and along $Z$ axis, respectively.
 $U_{ii}= {4\pi a_{ii} \hbar^2}/{m_i}$ and $U_{ij}={2\pi a_{ij} \hbar^2}(m_i+m_j)/{m_im_j}$ are the intra-component and the inter-component coupling strengths, respectively.  We consider an atomic  valance electron  of mass $m_e$  is  moving around the mean field of  core electrons and nucleus with total charge $+e$ and mass $m_n$. The CM coordinate with respect to laboratory frame  is $\textbf{R}=(m_e \textbf{r}_e + m_n \textbf{r}_n)/m_t $, where $m_t=m_e+m_n$ being the total mass.    Here $\textbf{r}_e$ and $\textbf{r}_n$ are the coordinates of the valance electron and the center of atom,  respectively. Therefore, the relative (internal) coordinate can be expressed as $\textbf{r}=\textbf{r}_e -\textbf{r}_n$.

As the BEC components are coupled to each other,  any perturbation to one of the components leads to the change in the CM wavefunction of the other component. Here, we consider the perturbation is coming from the interaction of the non-paraxial LG beam, which is  produced from  a circularly polarized paraxial pulse with OAM by passing  it through a lens with high numerical aperture (NA). The spot size of the paraxial LG beam  is  overfilled the entrance aperture radius of the objective to take full advantage of the high numerical aperture. Due to the diffraction from the edges of the objective and the focusing from the NA, the SAM and OAM of the light get coupled and form a superposition of plane waves having an infinite number of spatial harmonics \cite{Richards1959, Boivin1965}.  For the non-paraxial circularly polarized LG beam, the x, y, z-polarized component of the electric field \cite{Zhao2007, Bhowmik2018, Monteiro2009, Iketaki2007} in the laboratory coordinate system can be expressed   as

\begin{equation}
{E_x}(r^\prime,\phi^{\prime},z^\prime)=(-i)^{l+1}E_0(e^{il\phi ^\prime}I_0^{(l)}+e^{i(l+2\beta)\phi ^\prime}I_{2\beta}^{(l)}),
\end{equation}
\begin{equation}
{E_y}(r^\prime,\phi^{\prime},z^\prime)=\beta(-i)^{l}E_0(e^{il\phi ^\prime}I_0^{(l)}-e^{i(l+2\beta)\phi ^\prime}I_{2\beta}^{(l)}),
\end{equation}
\begin{equation}
{E_z}(r^\prime,\phi^{\prime},z^\prime)=-2\beta(-i)^{l}E_0e^{i(l+\beta)\phi ^\prime}I_{\beta}^{(l)},
\end{equation}
where $\beta$ is the polarization of light incident on the lens. Here, we consider that the light is  circularly polarized with  $\beta = \pm 1$.   The amplitude of the focused electric field is $E_0=\frac{\pi f}{\lambda} T_{o} E_{inc}$, where  $T_{o}$ is the objective transmission amplitude, $E_{inc}$ is the amplitude of incident electric field on the high NA lens and $f$ is the focal length related with $r^\prime$ by $r^\prime=f \sin\theta$ (Abbe sine condition). The coefficients $I_m^{(l)}$, where $m$ takes the values 0, $\pm1$, $\pm2$   in the above expressions, depend on focusing angle ($\theta_{max}$) by \cite{Zhao2007}
\begin{center}
\begin{equation}
\hspace{-2.1cm}I_m^{(l)}(r _\bot ^\prime ,z ^\prime)=\int_0^{\theta_{max}}d\theta\left({\frac{\sqrt{2}r_\bot^\prime }{w_0 \sin\theta}}\right)^{| l |}{(\sin\theta)}^{| l | +1} \sqrt{\cos\theta} g_{| m |}(\theta) J_{l+m}(kr_\bot^\prime \sin\theta)e^{ikz^\prime \cos\theta},
\end{equation}
\end{center}
where $r_\bot^\prime$ is the projection of \textbf{r$^\prime$} on the $xy$ plane, $w_0$ is the waist of  beam at the position of the objective entrance port and $J_{l+m}(kr_\bot^\prime \sin\theta)$ is cylindrical Bessel function. The angular functions are  $g_0 (\theta)=1+\cos\theta$, $g_1 (\theta)=\sin\theta$, and $g_2 (\theta)=1-\cos\theta$.

 Let $ \psi_i$ and $\Psi_i $ be the internal electronic  and the external CM wavefunctions, respectively, of $i$-th component of BEC. The total wavefunction of the  two-component BEC can be written as $\Upsilon(\textbf{R}_1,  \textbf{R}_2, \textbf{r}_1, \textbf{r}_2)=\Psi_1({\textbf{R}_1}) \Psi_2({\textbf{R}_2})\psi_1({\textbf{r}_1}) \psi_2({\textbf{r}_2})$. The atom-radiation interaction Hamiltonian, $H_{int}$,  is derived from the Power-Zienau-Wooley (PZW) scheme \cite{Babiker2002} which is beyond the level of dipole approximation.
 
 \begin{equation}
H_{int}=-\int d\textbf{r}^\prime P(\textbf{r}^\prime)\boldsymbol{.} \textbf{E}(\textbf{r}^\prime, t) +h.c.
\end{equation}
where $\textbf{E}(\textbf{r}^\prime, t)$ is the local electric field of the LG beam \ experienced by the atom. $P(\textbf{r}^\prime)$ is the electric polarization
given by $ P(\textbf{r}^\prime)=-e\frac{m_n}{m_t}\textbf{r}\int_0^1 d\lambda \delta \Big(\textbf{r}^\prime-\textbf{R}-\lambda\frac{m_n}{m_t}\textbf{r}\Big).$ If the LG beam (with OAM=$+l$ and SAM=$\pm1$) interacts with one of the components of the BEC (say, $n$-th), then the dipole transition matrix element  will be (for interaction with single-component BEC, see Ref.\cite{Bhowmik2016})

\begin{eqnarray}
\hspace{-2.2cm} M_{i \rightarrow f}^d & \hspace{-1.2cm}=&  \langle \Upsilon _f | H_{int} | \Upsilon _i \rangle  \nonumber \\
 &\hspace{-1.2cm}=& e\frac{m_c}{m_t}  \sqrt{\frac{8\pi}{3}}\Bigl[-\epsilon_{\pm 1}\langle \Psi _{nf}({\textbf{R}_n}) | I_0^{(l)}({\textbf{R}_n})e^{il\Phi} | \Psi _{ni}({\textbf{R}_n}) \rangle \langle \psi _{nf}({\textbf{r}_n}) | r Y_1^{\pm 1}(\boldsymbol{\hat{\textbf{r}}})| \psi _{ni}({\textbf{r}_n}) \rangle \nonumber \\
&\hspace{-1.2cm}-& \epsilon_{\mp 1} \langle  \Psi _{nf}({\textbf{R}_n}) | I_{\pm 2}^{(l)}({\textbf{R}_n})e^{i(l\pm 2)\Phi}| \Psi _{ni}({\textbf{R}_n}) \rangle          \langle \psi _{nf}({\textbf{r}_n})| r Y_1^{\mp 1}(\boldsymbol{\hat{\textbf{r}}})| \psi _{ni}({\textbf{r}_n}) \rangle  \nonumber\\ 
&\hspace{-1.2cm}\pm  &   \sqrt{2} i \epsilon_{0} \langle \Psi _{nf}({\textbf{R}_n}) |  I_{\pm 1}^{(l)}({\textbf{R}_n})e^{i(l\pm 1)\Phi} | \Psi _{ni}({\textbf{R}_n}) \rangle \langle \psi _{nf}({\textbf{r}_n}) | r Y_1^{0}{(\boldsymbol{\hat{\textbf{r}}}})| \psi _{ni}({\textbf{r}_n}) \rangle \Bigr]\nonumber \\
&\hspace{-1.2cm}\times & \prod_{p\neq n}\langle \Psi _{pf}({\textbf{R}_p}) | \Psi _{pi}({\textbf{R}_p}) \rangle \langle \psi _{pf}({\textbf{r}_p}) | \psi _{pi}({\textbf{r}_p}) \rangle,
\end{eqnarray}
where $\epsilon_\pm= (E_x \pm iE_y)/\sqrt{2}$ and  $\epsilon_0=E_z$. Eqn. (7) shows three possible hyperfine sub-levels of electronic transitions and this
part of the transition matrix element is calculated using  well known  relativistic coupled-cluster
theory \cite{Bhowmik2017a, Bhowmik2017b, Das2018, Biswas2018}. If  the interaction happened with paraxial LG beam, only one of the electronic transitions, corresponds to the first terms in the square bracket, would obtain depending on the choice of SAM of the paraxial LG beam. When a circularly polarized LG beam is focused, it creates different types of LG photon with three different local polarizations, generating three different electronic transitions. To conserve the total angular momentum of each photon, the  three different OAMs ($l$, $l\pm2\beta$, $l\pm\beta$) of field are transferred to the CM of the atoms of the interacting component of BEC. Since the motions of the two component are coupled, the generation of three different vorticities in one of the components of BEC, modifies the CM wavefunction of the other component in three different ways. Further, the interaction Hamiltonian  also depends on the focusing angle of the LG beam. Therefore, tuning of focusing angle of the LG beam directly affects the strength of interaction of the LG beam with BEC. Since, the coupling between the OAM and SAM of the focused LG beam creates  special kind of intensity distribution due to the non-vanishing contributions of Z-component \cite{Monteiro2009}, we expect longitudinal variation of Rabi frequencies during the interaction among the components of BEC.

\begin{figure}[!h]

\centering
\includegraphics[trim={0.5cm 0.5cm 1cm 1cm},width=13cm]{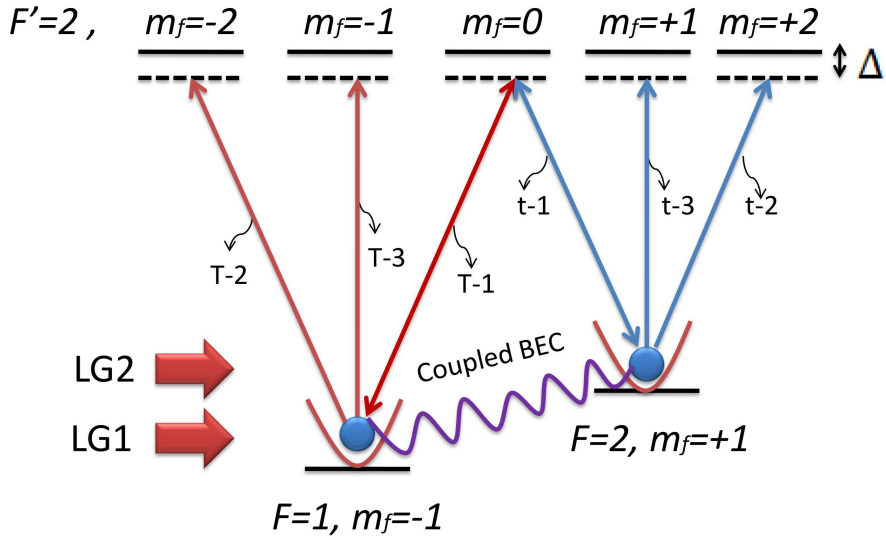}
\caption{Energy level scheme of the two-photon transitions at two-component BEC. Focused LG1 (OAM$=-1$ and SAM$=+1$) and LG2 (OAM$=+1$ and SAM$=-1$) beams are co-propagating and they interact with 1st and 2nd components of BEC, respectively.  The  ground states of 1st component and 2nd component of the  $^{87}$Rb BEC are $|  5s_{\frac{1}{2}} F=1, m_f =-1 \rangle$ and $| 5s_{\frac{1}{2}} F=2, m_f =+1 \rangle$, respectively. 
 $\Delta=-1.5$ GHz represents two-photon detuning. T-1, T-2, T-3, t-1, t-2, and t-3 are Rabi frequency of two-photon transitions.}
\end{figure}

\section{NUMERICAL RESULTS AND INTERPRETATION}

We consider  co-propagating  two  sets of beams, say LG1 and LG2. Each set contains one LG  and  one Gaussian beams. Individual components of the coupled $ ^{87} $Rb BEC are considered to be non-rotating. The BEC is  prepared in  a harmonic potential using the two hyperfine states $\psi_1=| 5S_{\frac{1}{2}}, F=1, m_f =-1 \rangle$ and $\psi_2=| 5S_{\frac{1}{2}}, F=2, m_f =1 \rangle$ (see FIG. 1). They are designated, henceforth, as BEC-1 and BEC-2, respectively. The  interaction of the focused LG beam  with the individual  components of the BEC produces three angular momentum channels \cite{Bhowmik2016}.  According to the Eq. (7), these  three angular momentum channels  generate   +1, $-1$ and 0 units of topological charge at the CM of the atom of  BEC. The proper choice of polarizations of the Gaussian beams  can Raman-excite the atoms to the different stoke or anti-stoke electronic states.  Let us name the channels as T-1, T-2, and T-3, respectively, for  BEC-1, and t-1, t-2, and t-3, respectively, for BEC-2. These three angular momentum channels correspond to  the different Raman electronic transitions through the different intermediate states. For  BEC-1, the channels have three intermediate   electronic hyperfine states,  $| 5p_{\frac{3}{2}}, F=2, m_f =0 \rangle$, $| 5p_{\frac{3}{2}}, F=2, m_f =-2 \rangle$, and $| 5p_{\frac{3}{2}}, F=2, m_f =-1 \rangle$, respectively.   In case of BEC-2, the intermediate   electronic hyperfine states are   $| 5p_{\frac{3}{2}}, F=2, m_f =+2 \rangle$, $| 5p_{\frac{3}{2}}, F=2, m_f =0 \rangle$,  and $| 5p_{\frac{3}{2}}, F=2, m_f =+1 \rangle$, respectively. Depending on the requirement of the problem, we can choose a particular Gaussian beam for the channel of our interest. Atoms excited by other channels  will be lost from the trap due to linear momentum transferred from the focused LG beam. 

A brief discussion on  the  interaction of the two-component BEC  with the paraxial LG beams will be useful before considering non-paraxial beam in the study. However, the detailed structure of the above BEC mixture for different inter-component interactions and number of atoms along with the formalism of the interaction with paraxial LG beam are available in our recent  paper \cite{Anal2018}. The asymmetry parameter of the harmonic trap is $\lambda _{tr} =\omega _Z /\omega _\bot =2$ with the axial frequency is $\omega _Z /2\pi =40$ Hz. The characteristic length is  $a _\bot =4.673$  $\mu$m.   The intensity of the paraxial LG beam is considered, $I =10^2 $ W cm$^{-2}$ but the intensity of the non-paraxial LG beam before focusing is assumed 10 mW m$^{-2}$ and its waist $w _0 =10 ^{-4}$ m. The intra-component $s$-wave scattering lengths are $a_{11}=1.03\times 5.5$nm, $a_{22}=0.97\times 5.5$nm \cite{Hall1998} and inter-component $s$-wave scattering length  $a_{12}=a_{21}=$ $g\times 5.5$ nm, where $g$ is a parameter which can be tuned using Feshbach resonance \cite{Chin2010, Inouye1998}.   For simplicity, we  consider that both the hyperfine states are populated by   equal number of atoms. 

\subsection{Density structure of non-vortex two-component BEC revisited}

FIG. 2 presents the initial non-vortex density distribution of the two-component BEC at $z=0$ plane for $N=10^6$.  FIG 2(a) to 2(i) represent the distribution with  increasing  inter-BEC coupling strength, $g$. Since BEC-1 has  relatively stronger intra-BEC strength compared to BEC-2,  BEC-1 is radially more expanded than BEC-2 at $g=0$.  However, this yields relatively less  central density of BEC-1  compared to BEC-2 (see FIG 2(a)). As the mutual interaction between the components of BEC is increased,   the components start departing from each other (FIG. 2(b) and 2(c)). Eventually after a certain value of $g$, a part of the BEC-2 breaks and grows at the outer region of BEC-1 (FIG 2(d), 2(e), 2(f) and 2(g)). Further increase of inter-component interaction even breaks BEC-1 in some parts and it appears at the surrounding of BEC-2 (FIG. 2(h) and 2(i)). Therefore,    multi-ring shaped density profiles are obtained  in the x-y plane with increasing $g$ value. 

\begin{figure*}[!h]
	\begin{center}	
\subfloat[]{\includegraphics[trim = 13cm 1.5cm 7cm 2cm,scale=.16]{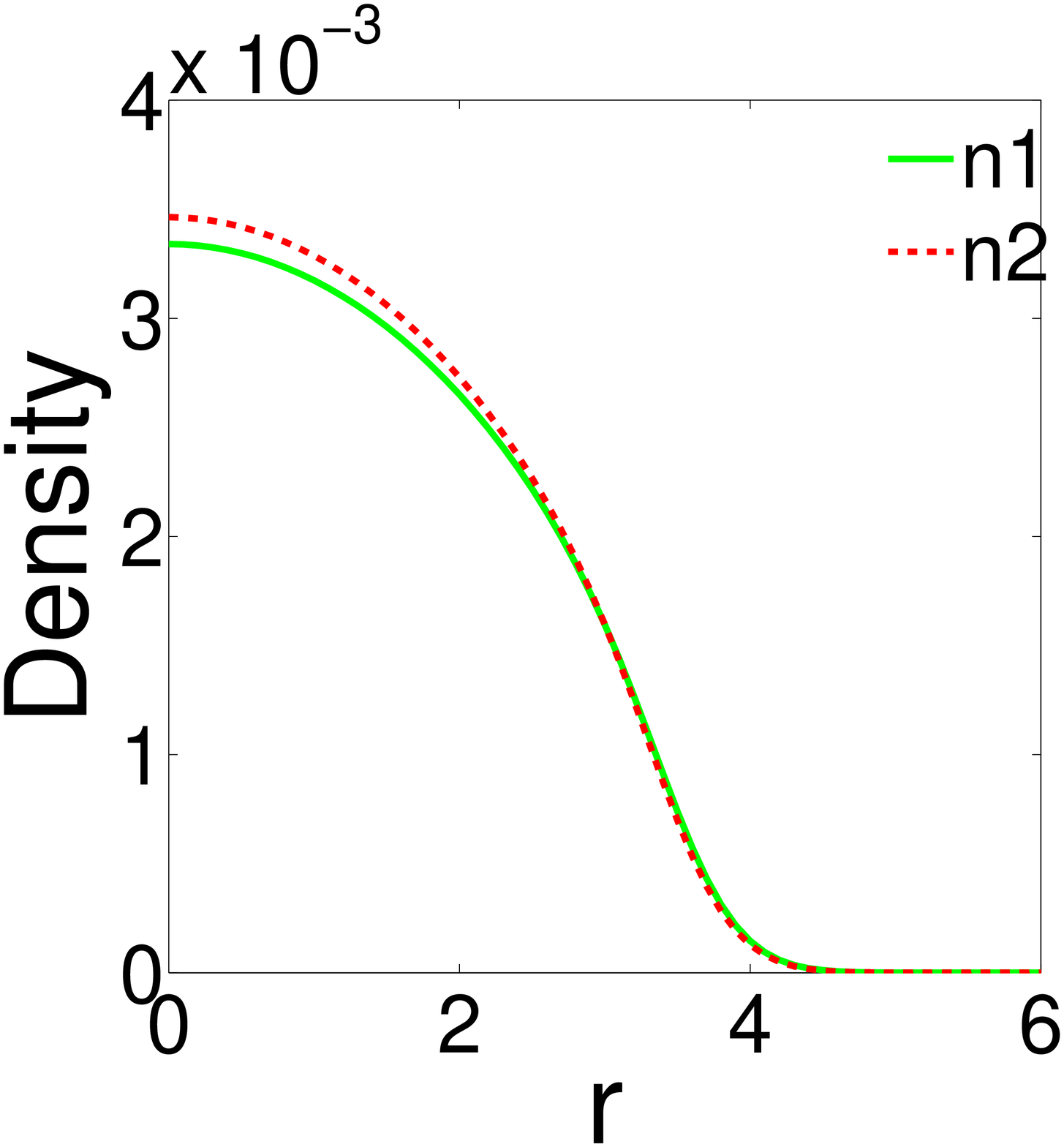}}
\subfloat[]{\includegraphics[trim = 9cm 1.5cm 7cm 0cm,scale=.16]{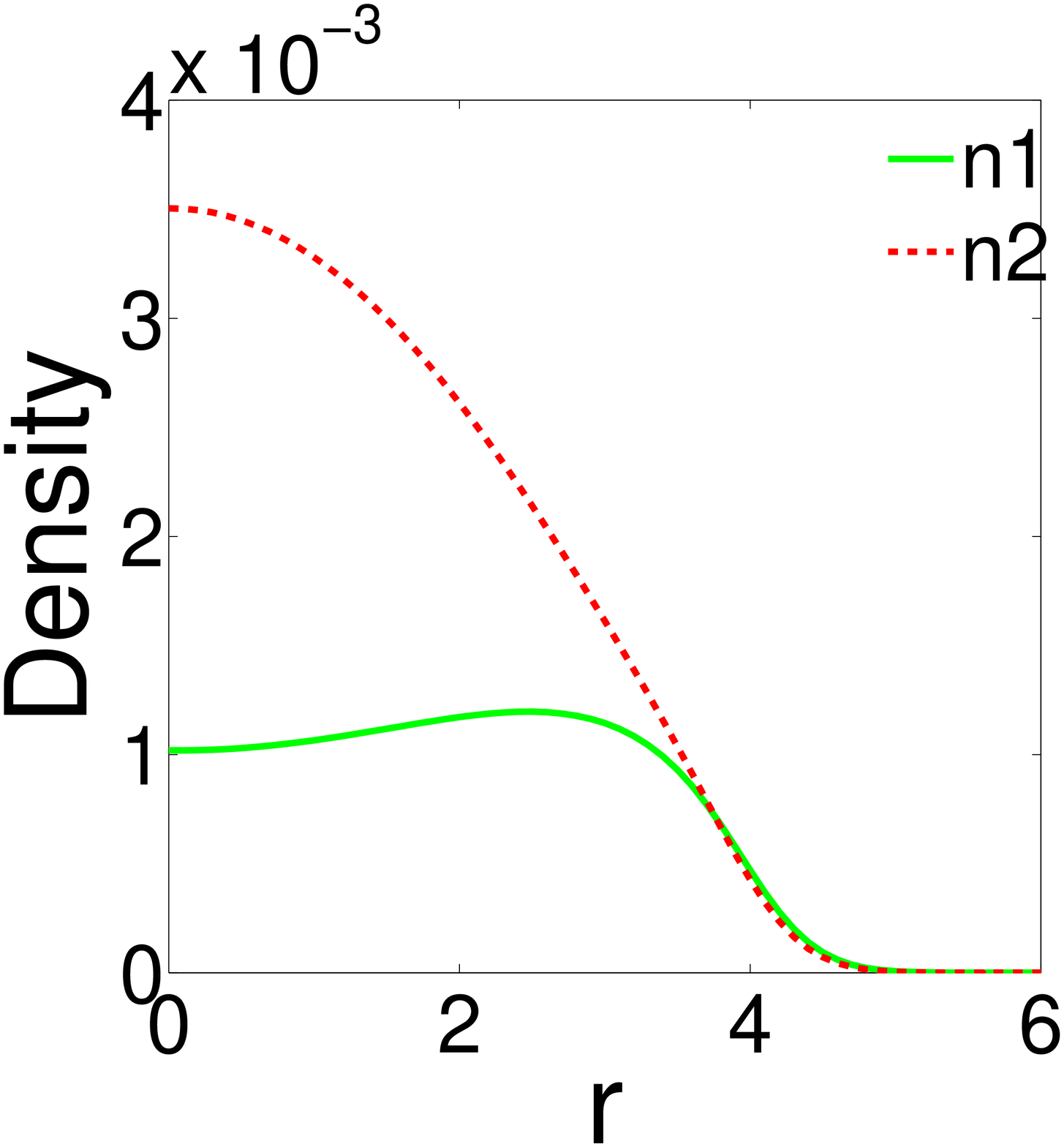}}
\subfloat[]{\includegraphics[trim = 7cm 1.5cm 7cm 2cm,scale=.16]{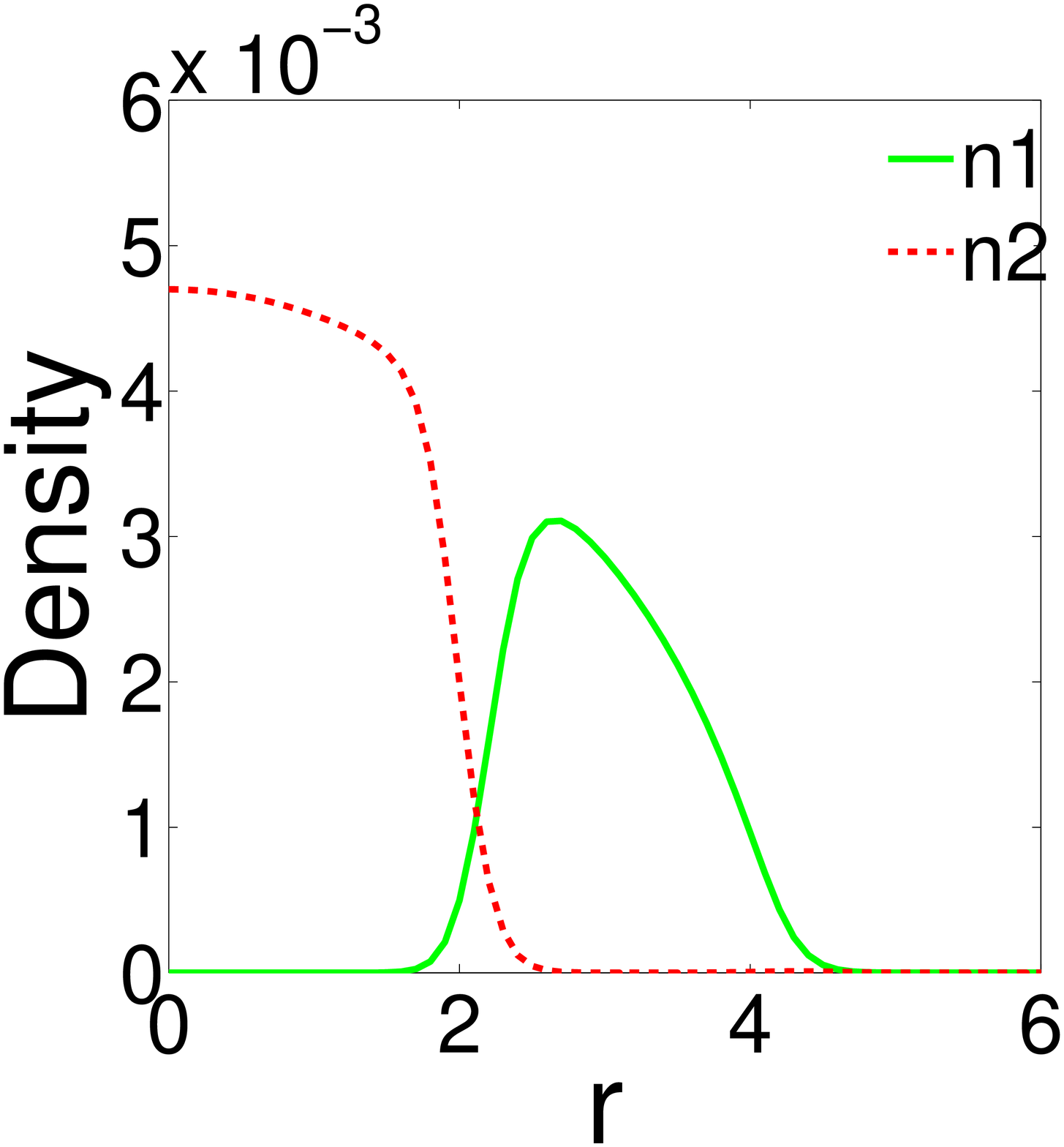}}\\
\subfloat[]{\includegraphics[trim = 13cm 1.5cm 7cm 2cm,scale=.16]{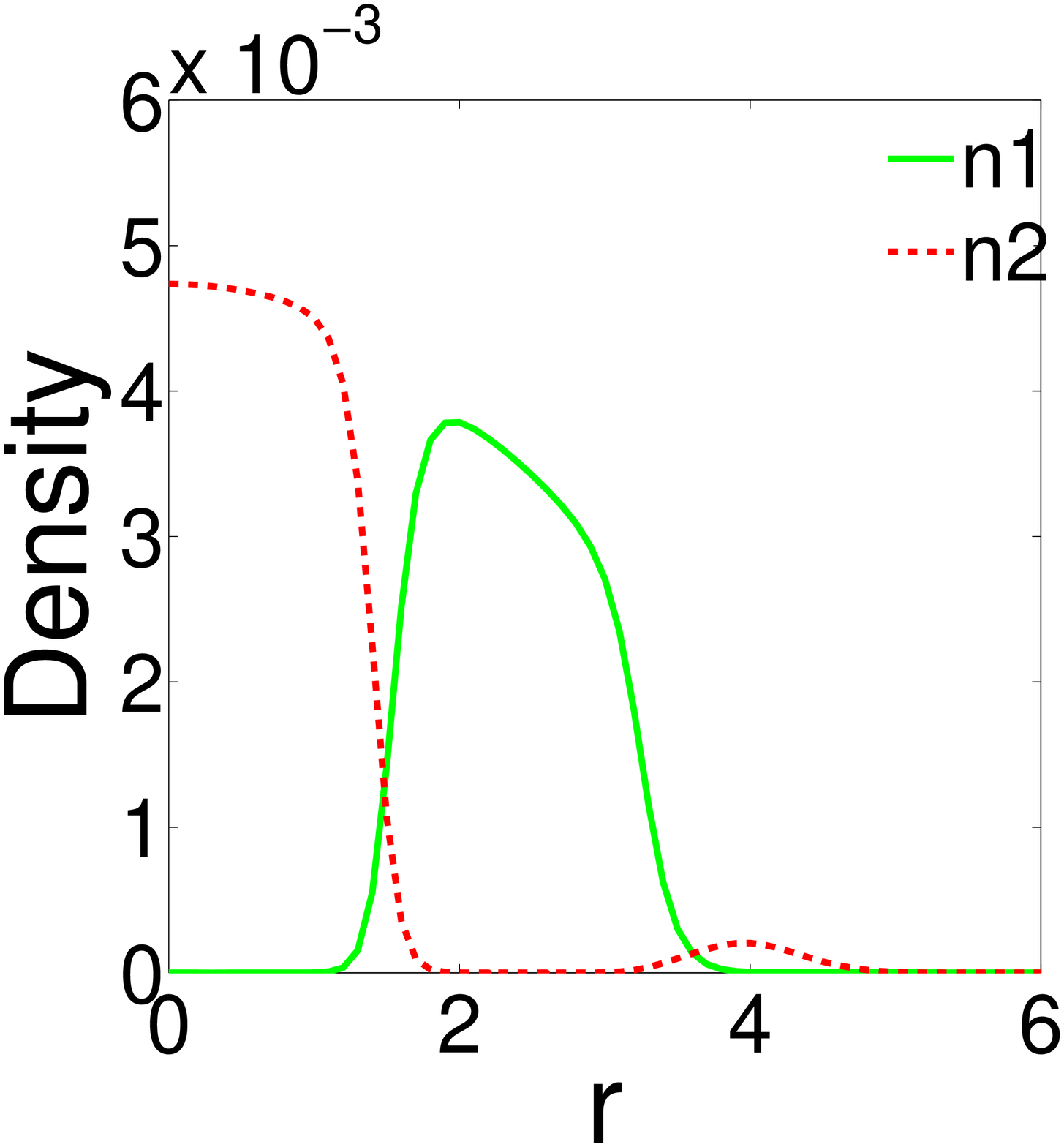}}
\subfloat[]{\includegraphics[trim = 9cm 1.5cm 7cm 2cm,scale=.16]{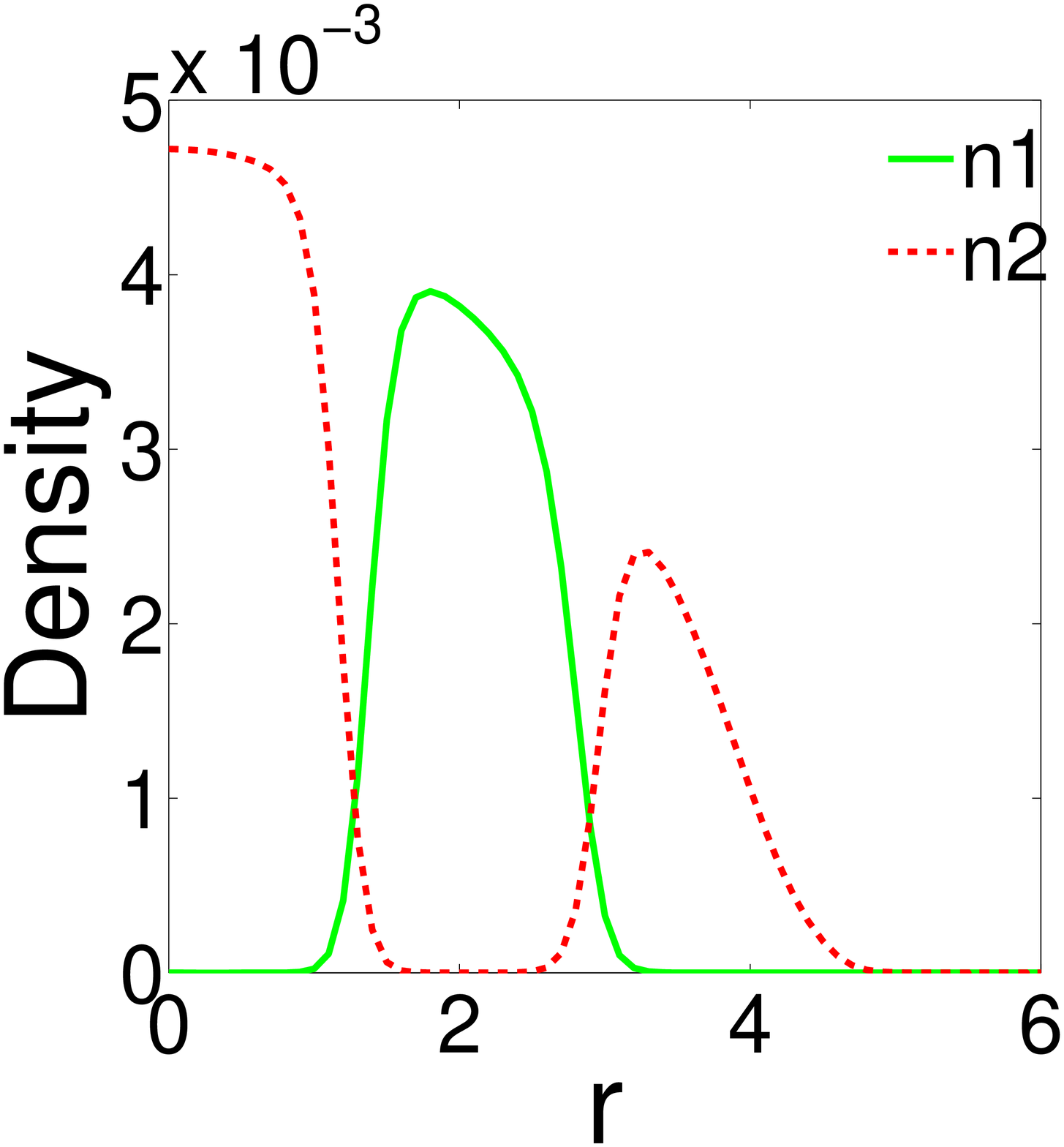}}
\subfloat[]{\includegraphics[trim = 7cm 1.5cm 7cm 0cm,scale=.16]{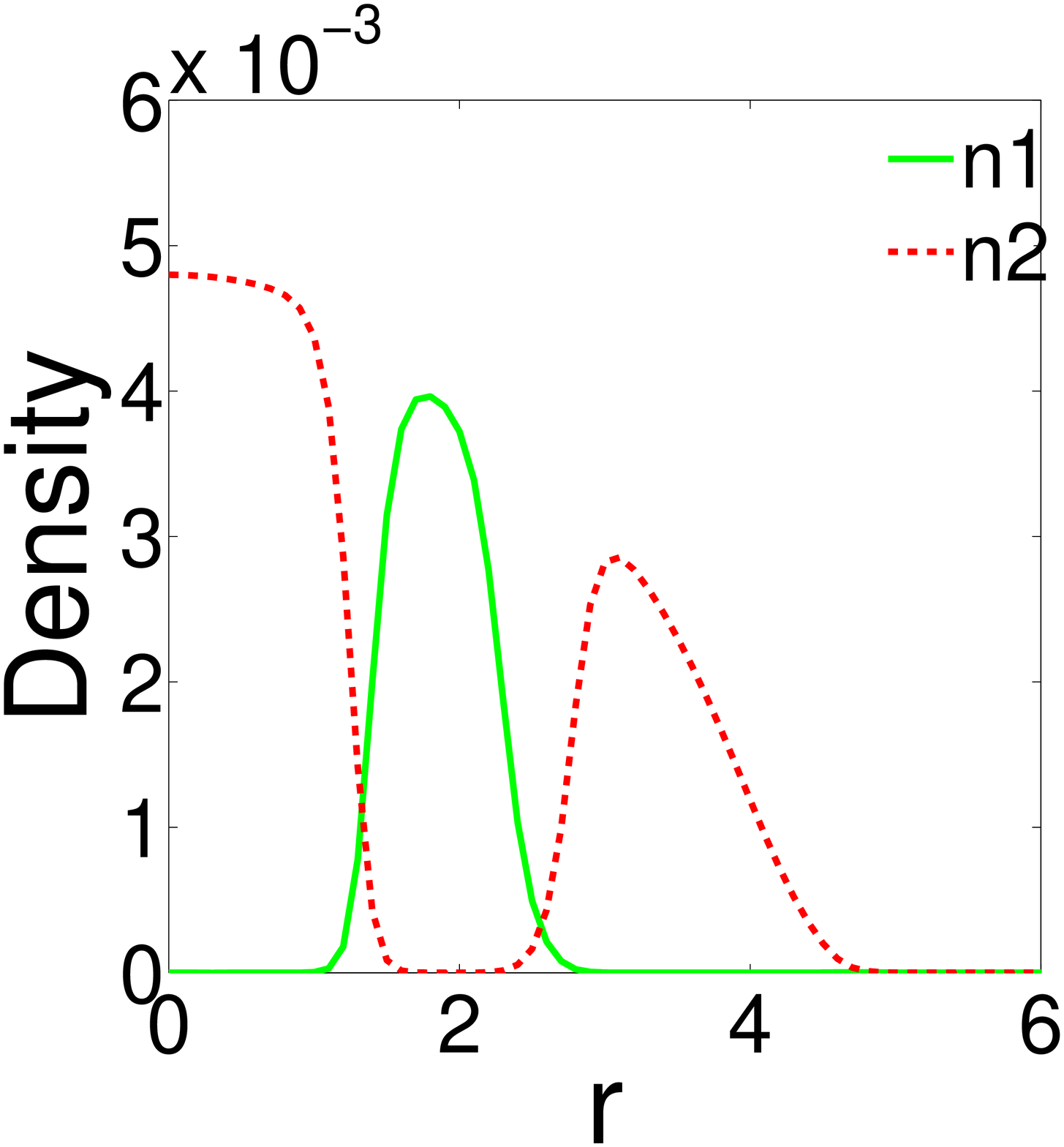}}\\
\subfloat[]{\includegraphics[trim = 13cm 1.5cm 7cm 2cm,scale=.16]{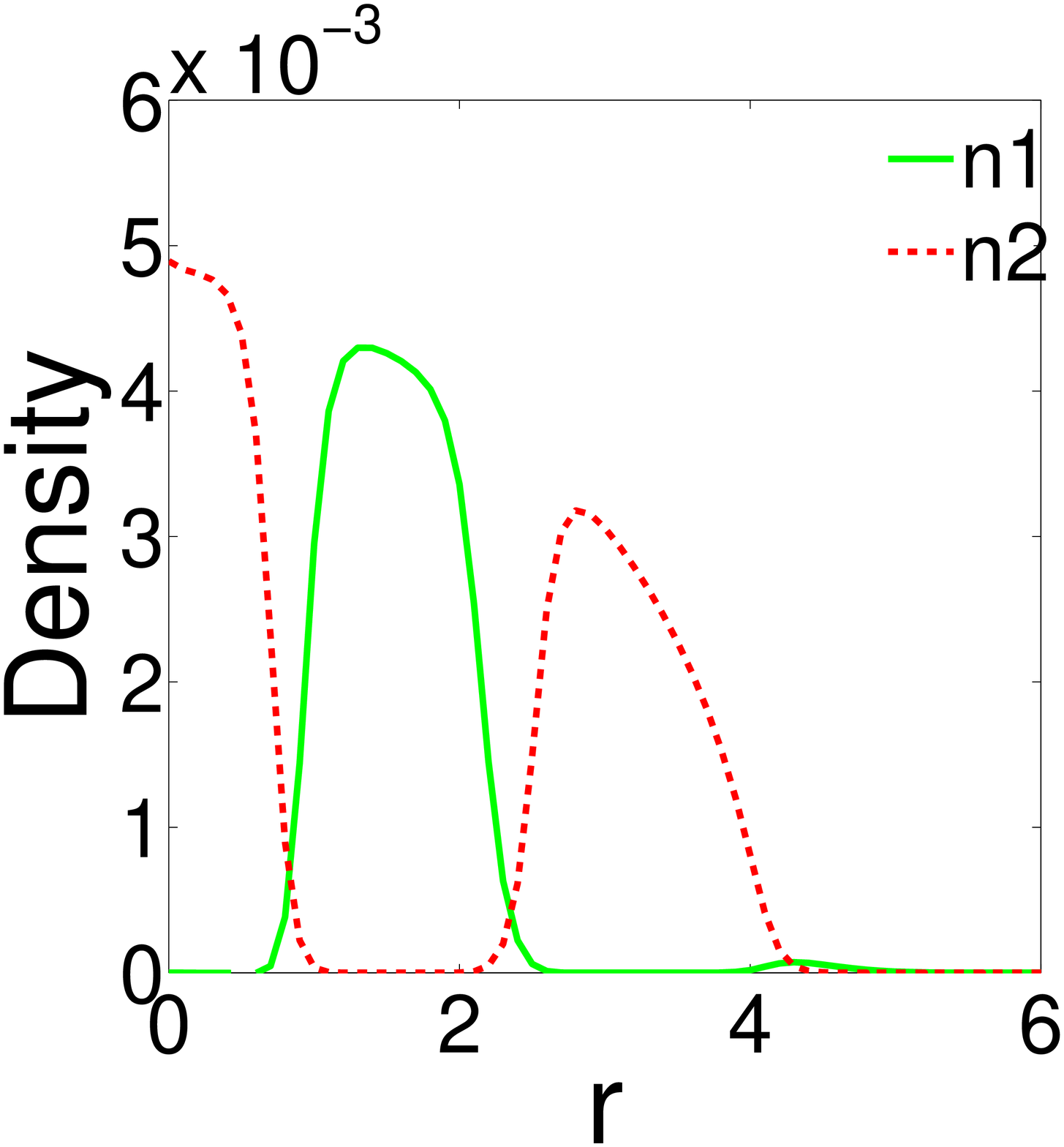}}
\subfloat[]{\includegraphics[trim = 9cm 1.5cm 7cm 0cm,scale=.16]{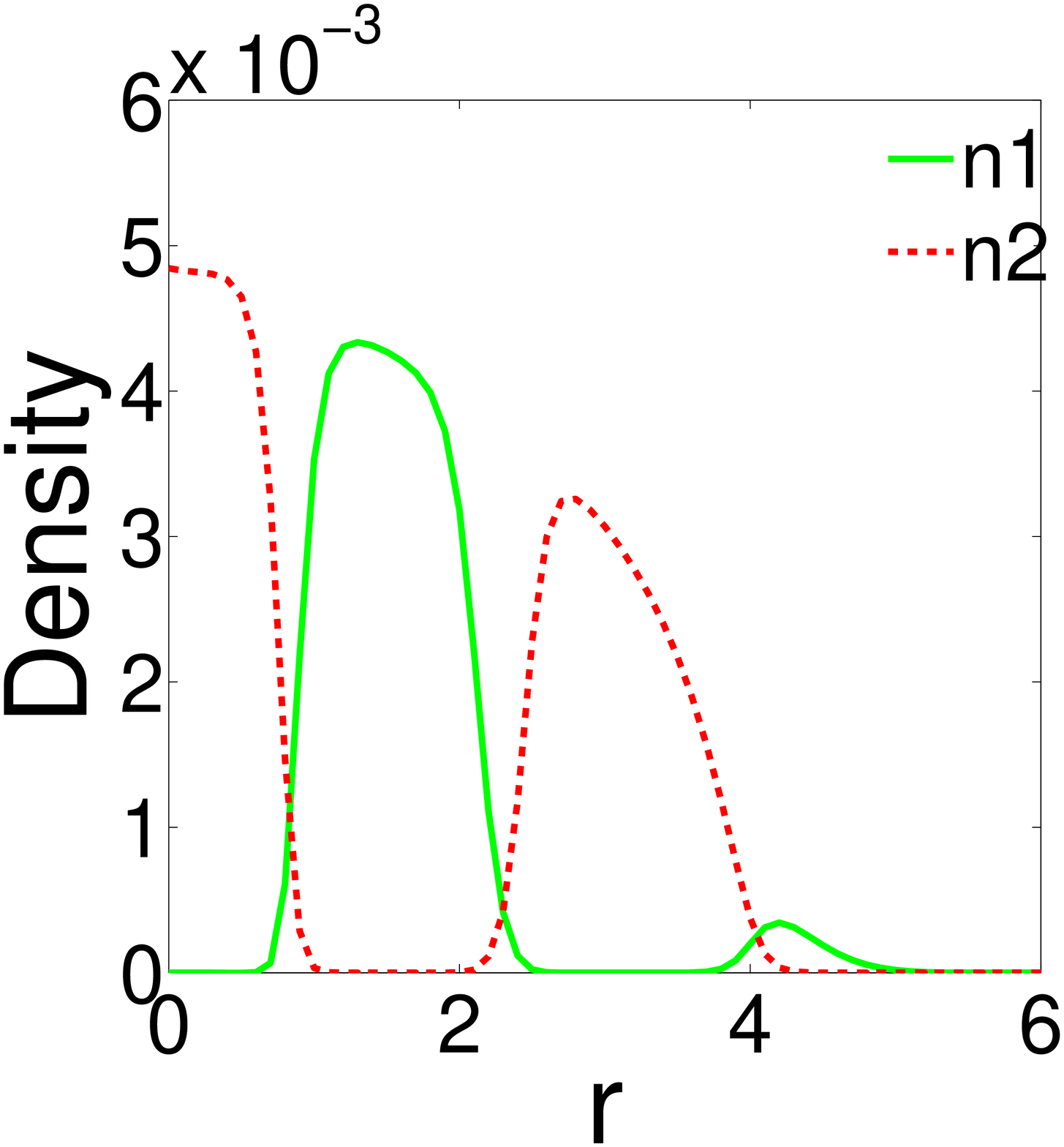}}
\subfloat[]{\includegraphics[trim = 7cm 1.5cm 7cm 0cm,scale=.16]{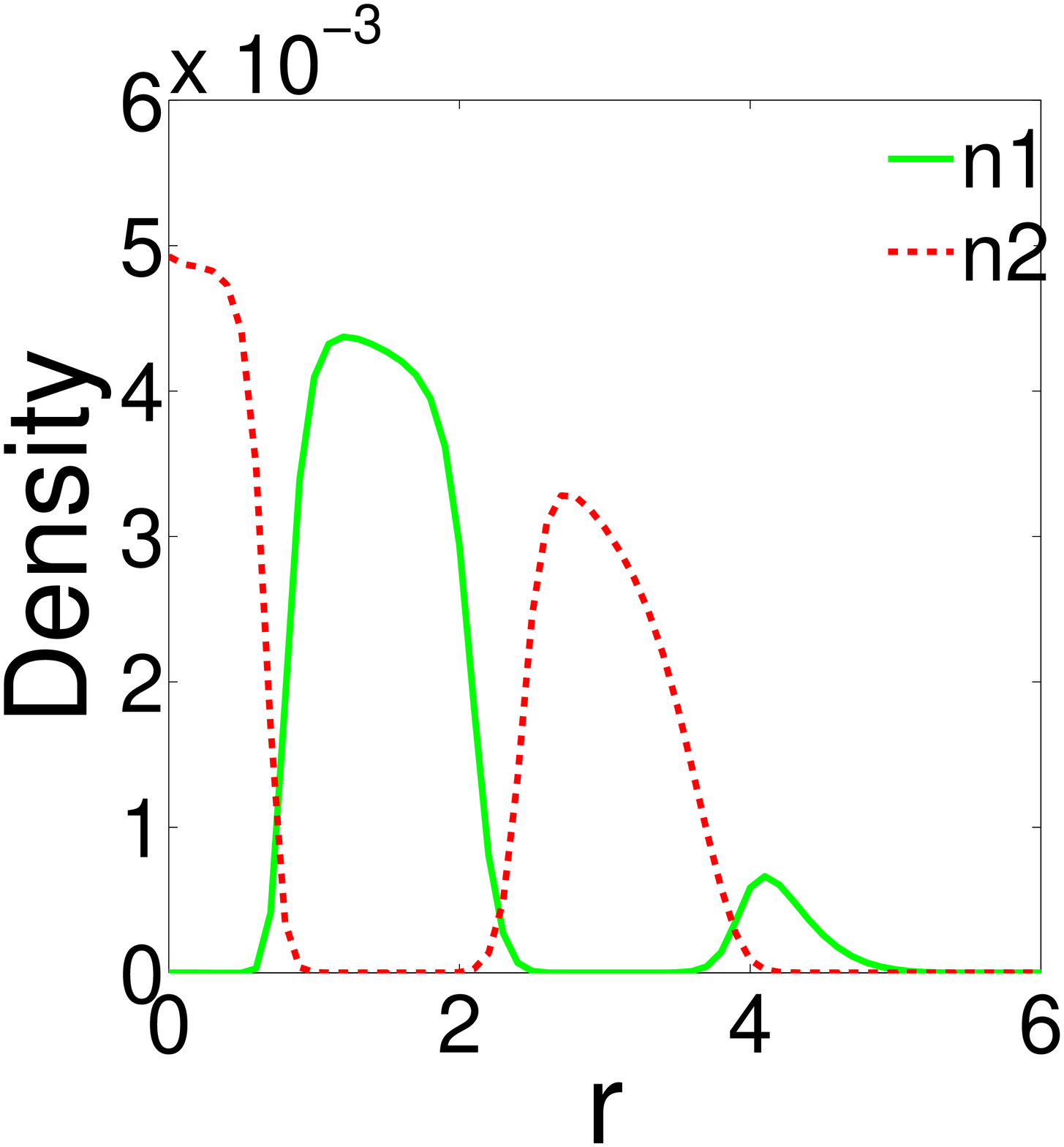}}
\caption{Plots of the density (in unit of $a_\perp^{-3}$ ) of non-vortex BEC-1 (solid lines, n1) and BEC-2 (dotted line, n2) with respect to distance from the trap axis are presented for N=$10^6$. Figure (a) $g=0$, (b) $g=0.50$, (c) $g=0.80$, (d) $g=1.00$, (e) $ g=1.10$, (f) $ g=1.30$, (g) $g=1.60$, (h) $g=1.70$, and (i) $ g=1.80$. r is in the unit of the harmonic oscillator length scale $(a_\perp)$.}
\end{center}
\end{figure*}

\subsection{Interaction with the paraxial LG beams}
 Let us consider that the paraxial LG1 (with OAM$=-1$ and SAM$=+1$) and LG2 (with OAM$=+1$ and SAM$=-1$) beams  are  impinged on BEC-1 and BEC-2, respectively,  as shown in FIG. 1. Therefore, for both the components of the BEC, two photon Raman transitions  are performed with  co-propagating LG and Gaussian (G) beams based on dipole transitions.   Due to the particular selection of OAM and SAM  of the paraxial LG beams, only T-1 and t-1  channels will be available with one kind of Gaussian beam. The channels transfer $-1$ and $+1$ units of  OAM to the atoms, respectively, to a particular electronic state, say,  $\psi_2=| 5S_{\frac{1}{2}}, F=2, m_f =1 \rangle$.  Therefore,   superposition of the vortex and antivortex states is created  at the  center of mass of the condensate component  corresponding to the electronic state, i.e., BEC-2. To examine the effect of inter-component coupling to the T-1 and t-1 transition channels, we have studied and analyzed  corresponding two-photon Rabi frequencies in FIG. 3(a).  The above non-vortex density profiles are considered as initial wavefunctions of the two-components BEC. In this dipole transition, the OAM of light does not contribute to the internal motion of the atoms. Therefore, the Rabi frequencies are calculated from the multiplications of  the center of mass matrix elements  and the electronic matrix elements involving OAM and SAM, respectively. This has been discussed in detail in our earlier paper \cite{Anal2018} with many distinct physical features. 
 
 \begin{figure*}[!h]
 	\subfloat[]{\includegraphics[trim =  1cm 0.5cm 0.1cm 0.1cm,scale=.30]{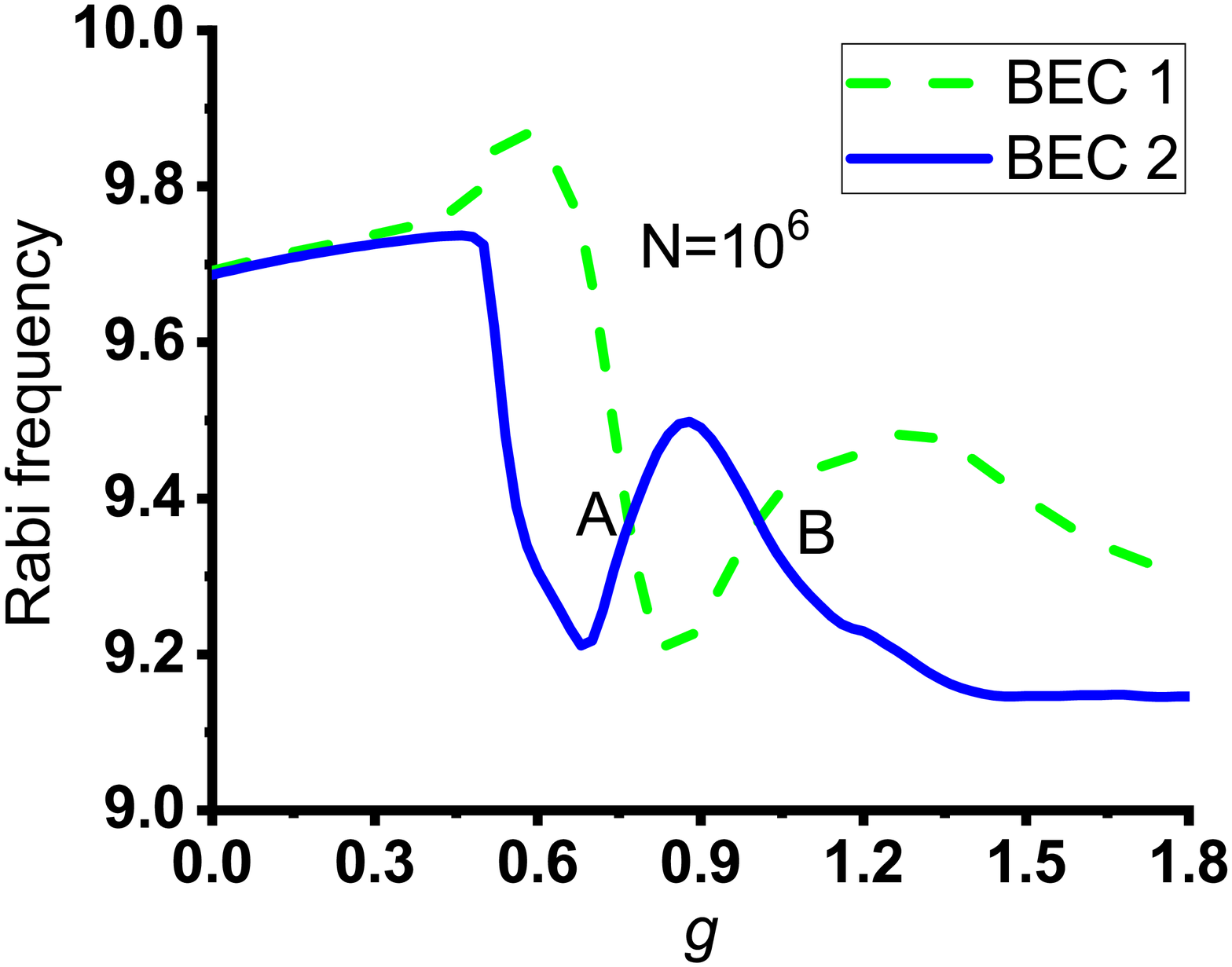}}
 	\subfloat[]{\includegraphics[trim = 1cm 0.5cm 0.1cm 0.1cm, scale =.30]{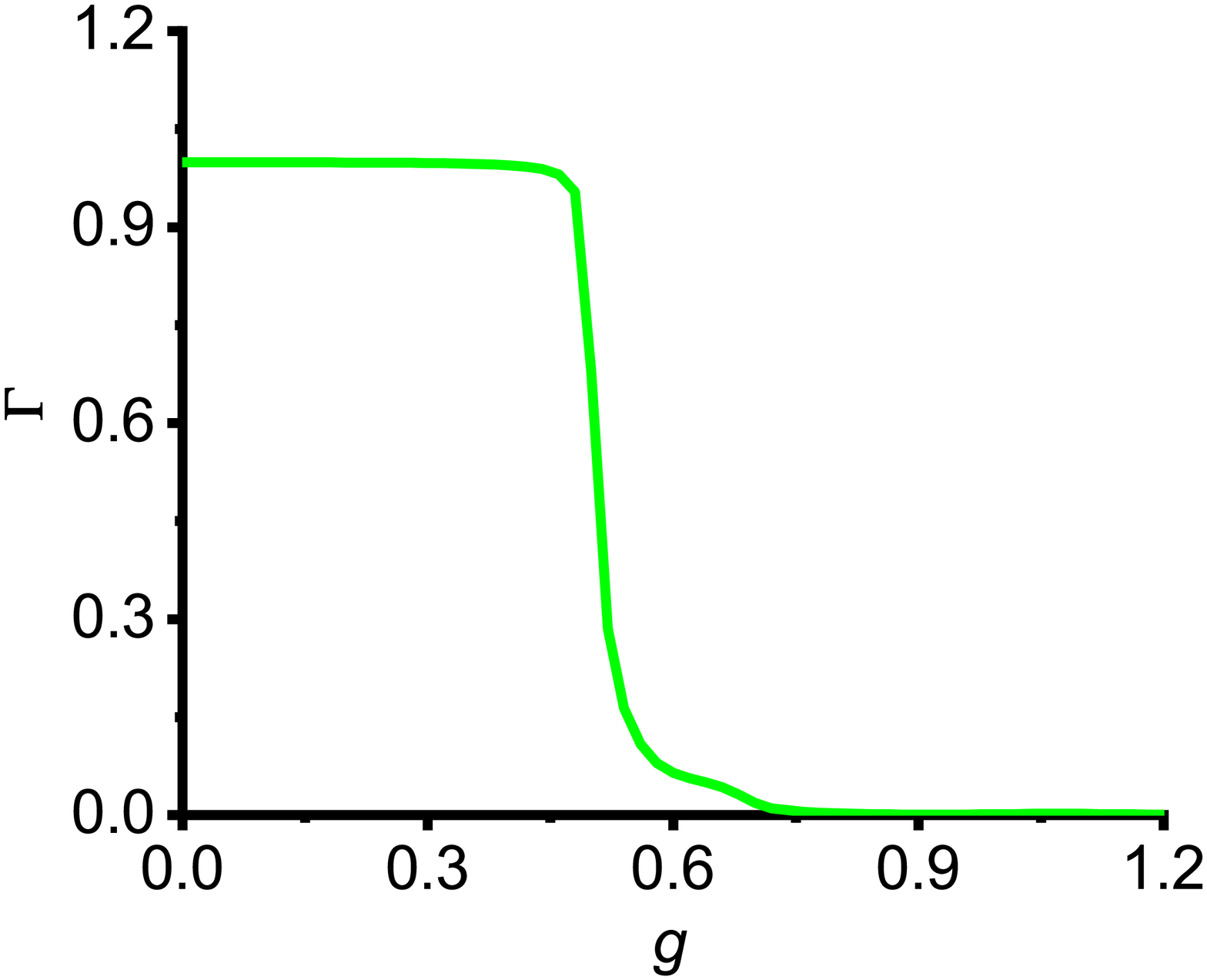}}
 	\caption{(a) Variations of dipole Rabi frequency (in Sec$^{-1}$) for T-1 and t-1 channels of Figure 1 (for paraxial laser) are plotted with respect to inter-component interaction strength $(g)$ on a semi-log scale. Initial states of both the components of BECs are considered as non-vortex. b) Variation of overlap matrix $(\Gamma)$ between the non-vortex condensed components is ploted with respect to $g$-values}
 \end{figure*}
 
 The Rabi frequency profiles of T-1 and t-1 transition channels are presented in FIG. 3(a) considering  both LG1 and LG2 as  paraxial. The figure interprets  that  BEC-1 and BEC-2 have almost the same initial density structures for $g$-values between 0 and 0.4.  Around $g=0.6$, the local peak of Rabi frequency variation for BEC-1 in contrast to the constant descending  profile  of BEC-2 can be explained clearly from their initial density structures  around that coupling region. Initial density profile at that $g$-value shows that BEC-2 is compressed around the center of the trap and BEC-1 is moved away from that center. There is a coincident observed  near $g=0.9$. There the Rabi frequencies  becomes locally minimum for BEC-1, but  locally maximal for BEC-2. Also it is the $g$-value at which the initial BEC components becomes totally immisible as seen from FIG. 3(b) containing the plot for variation of overlap between the components. The overlap parameter is calculated using Eq. (8) of reference \cite{Jain2011}. 

\begin{equation}
\Gamma=\frac{[\int n1(r)n2(r)dr]^2}{[\int n1(r)^2 dr][\int n2(r)^2 dr]}.
\end{equation}
 The complete overlap, i.e.  $ \Gamma=1 $ indicates
total mixing between the components, whereas for complete phase separation condition we have $ \Gamma=0 $. The cutting points A and B of the Rabi frequency distributions in FIG. 3(a)  describe the population of the vortex and antivortex states at BEC-2 will be same.  Therefore, these inter-component coupling strengths are ideal for a maximally coherent  fringe pattern of interference. When the components are non-miscible, we may get interesting vortex-dipole dynamics \cite{Sang2017} which is beyond the scope of discussion of this paper.

However, the interactions of the non-paraxial or focused LG beam with the two-component BEC will not only provide  enhanced Rabi frequencies due to increased intensity, but also will generate different channels of transitions along with their external control mechanism as  discussed in the following  subsection.
  
 \begin{figure*}[!h]
 	\subfloat[]{\includegraphics[trim = 1cm 0.5cm 0.1cm 1.5cm, scale=.30]{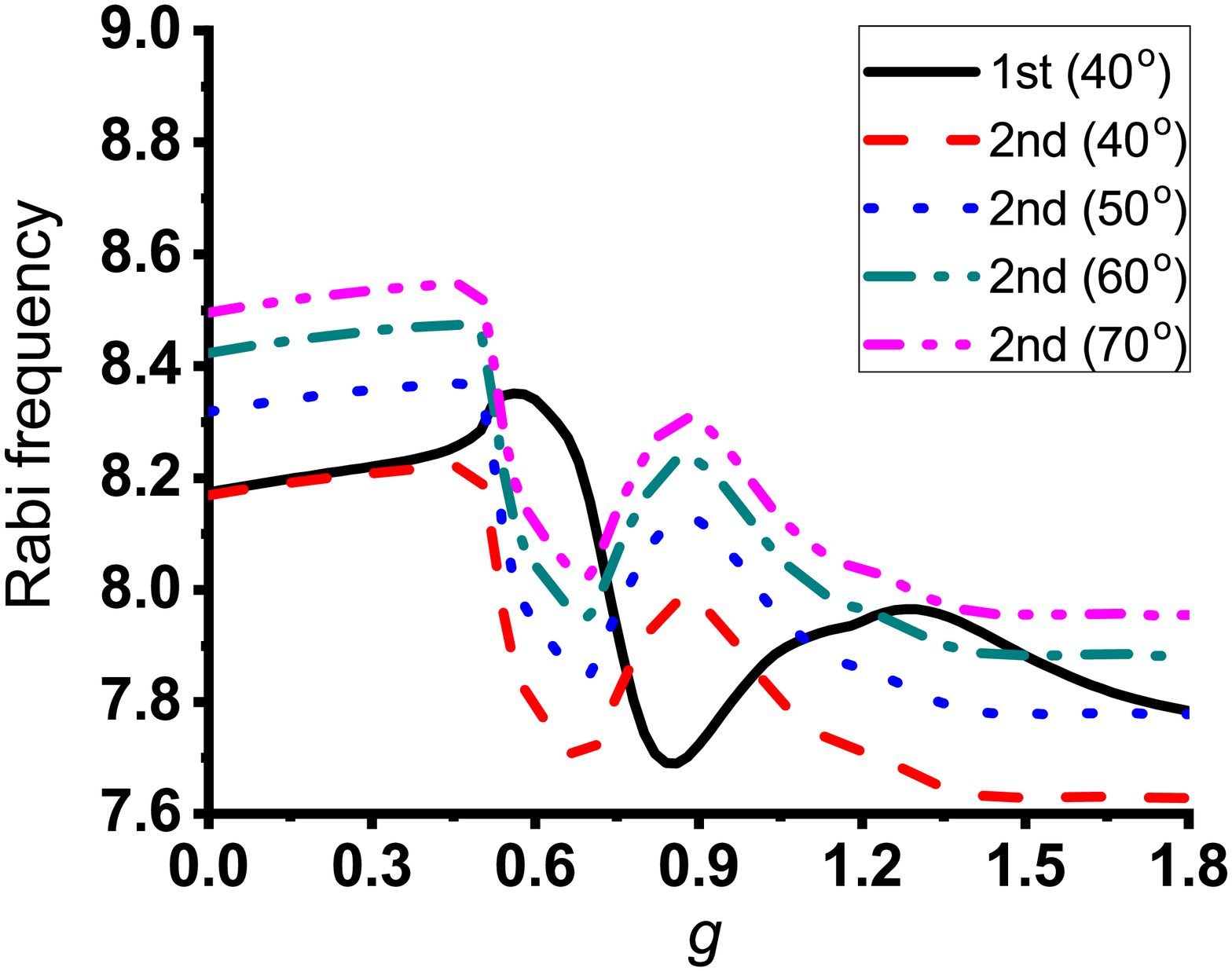}}
 	\subfloat[]{\includegraphics[trim = 1cm 0.5cm 0.1cm 1.5cm, scale=.30]{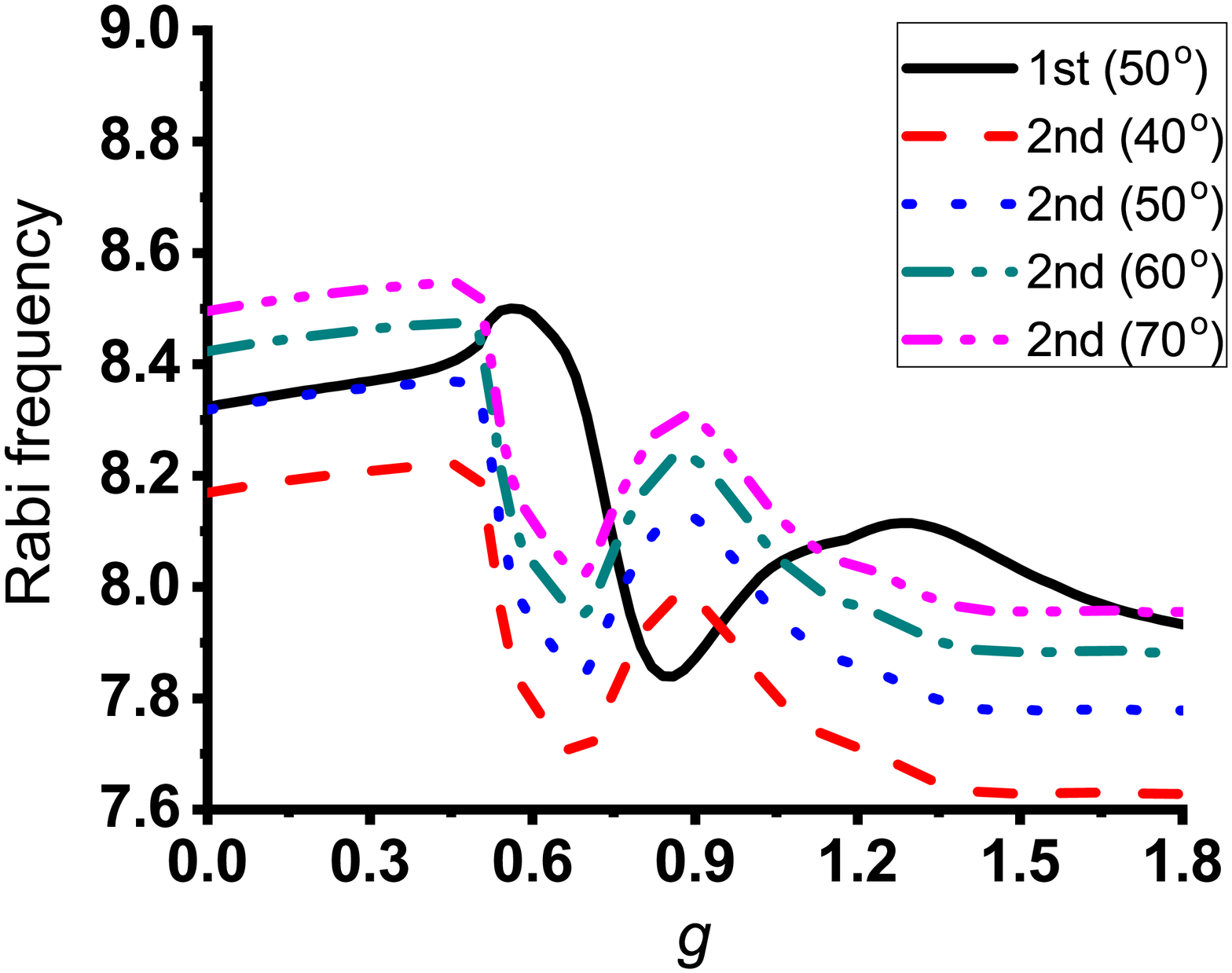}}\\
 	\subfloat[]{\includegraphics[trim =  1cm 0.5cm 0.1cm 0.1cm,scale=.30]{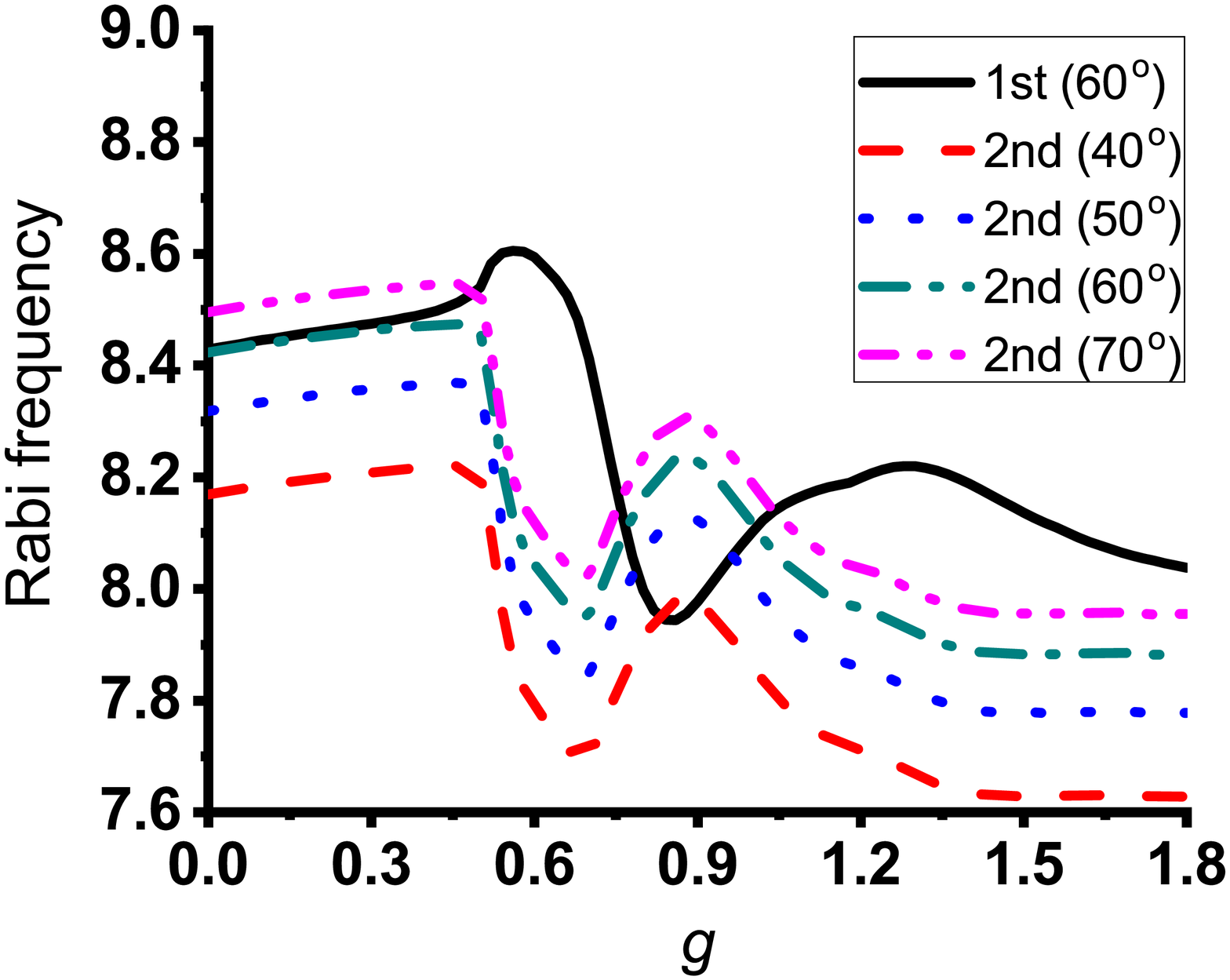}}
 	\subfloat[]{\includegraphics[trim =  1cm 0.5cm 0.1cm 0.1cm, scale=.30]{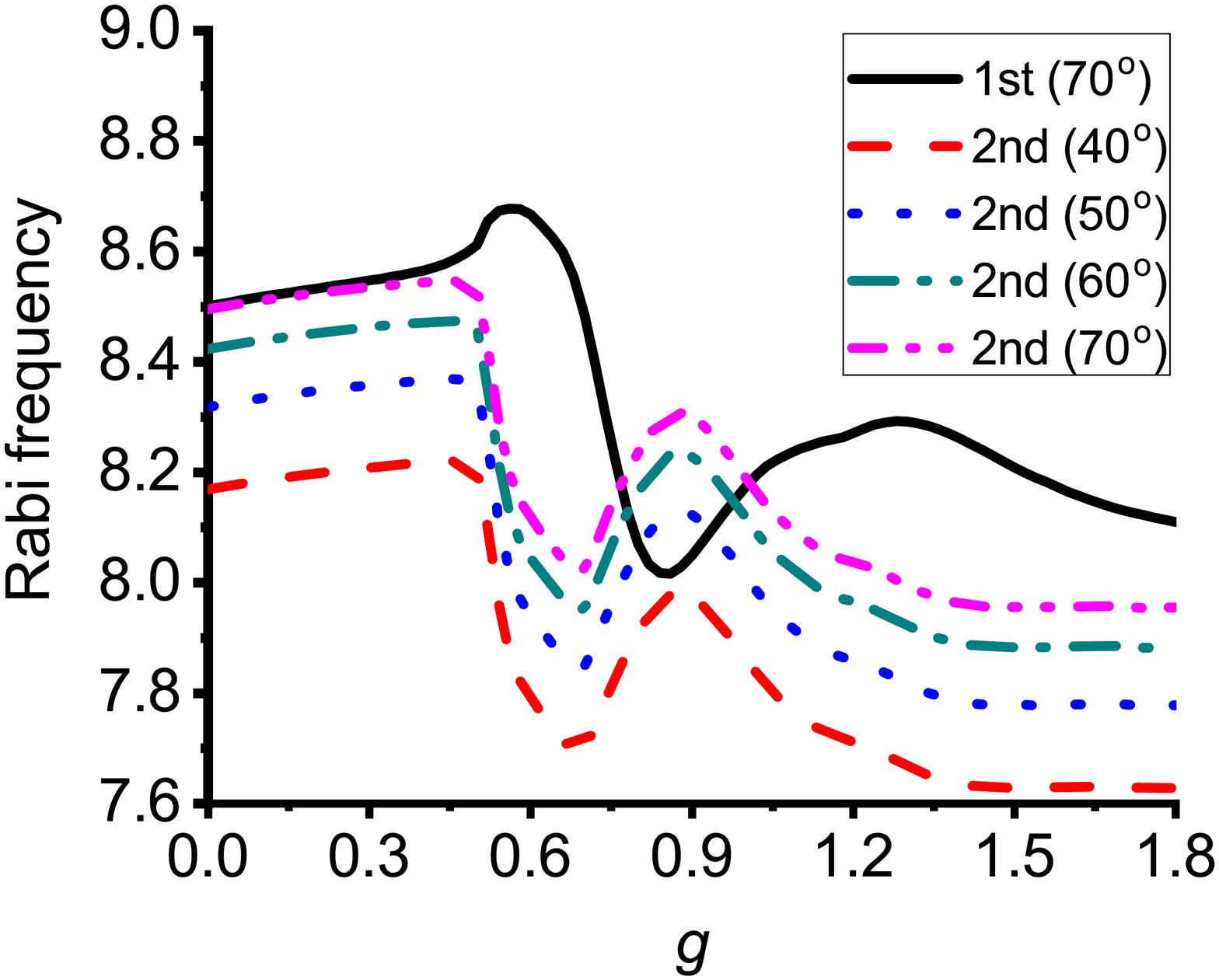}}
 	\caption{Variations of dipole Rabi frequency (in Sec$^{-1}$) of T-1 ("1st") and t-1 ("2nd") channels are plotted with  inter-component interaction strength $(g)$ on a semilog scale for different focusing angles of the LG beams. }
 \end{figure*}

\subsection{Interaction with the Non-Paraxial LG beams}
 
The interaction of the two-component BEC with the non-paraxial LG beams is the main theme of this work. The interactions open up different channels of transitions with variable strengths depending on the parameters of the LG beams.  Lets  chose first one of the dominant  options, where before focusing the OAM and SAM of LG1 (with OAM$=-1$ and SAM$=+1$) and LG2 (with OAM$=+1$ and SAM$=-1$) are such that the strengths of T-1 and t-1 transitions are the strongest among each set of transitions (see FIG. 1).  In fact, they  are the transitions which were involved in the above mentioned paraxial case. The difference is expected to observe in the strength of Rabi frequencies and the effect of the other transition channels through which atoms are lost from the trap.

 FIG. 4 shows the variation of the Rabi frequencies, calculated using expression Eq. (2.7),  with the inter-component interaction strengths, $g$ of the components of BEC at different focusing angles of the LG beam. The focusing angles of  LG1   are  considered 40$^{\circ}$, 50$^{\circ}$, 60$^{\circ}$ and 70$^{\circ}$ in FIG. 4(a), 4(b), 4(c) and 4(d) , respectively. An overall increase in the  Rabi frequencies is found with the increase in the focusing angle compared to the paraxial beam (compare with FIG. 3(a)).  This is understandable as  more number of photons are available for interaction with atoms trapped in the harmonic potential having a smaller cross-section compared to the paraxial beam size. In each of the plots of the FIG. 4, the focusing angles of  LG2 varies from 40$^{\circ}$ to 70$^{\circ}$ in order to analyze the mutual variations of the component wise interactions.  
 
 Unlike the paraxial case, the crossing points A and B of the Rabi frequency profiles for both the components  can be tuned  by changing the focusing angle of the light beam. It means that  the maximally coherent interference pattern can be achieved even at $g$-value away from the A and B points obtained in the paraxial case. In other words,  we will be able to estimate the focusing angles of LG1 and LG2 by observing the perfect interference pattern by tuning inter-component coupling strength of the BEC mixture. In certain combinations of focusing angles, particularly for a large difference in focusing angles between LG1 and LG2, the cutting points are not available due to the comparatively large enhancement of the average  Rabi frequency for the more  focused beam case (See black solid line and red dashed line in fig 4(d)).

\begin{figure}[!h]

\centering
\includegraphics[trim={0.5cm 0.5cm 1cm 1cm},width=13cm]{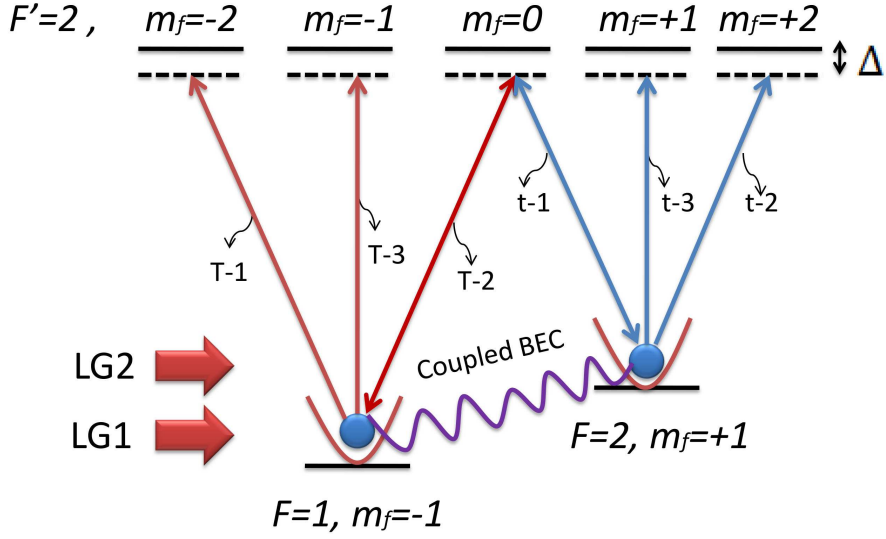}
\caption{Energy level scheme of the two-photon transitions. Focused LG1 (OAM$=+1$ and SAM$=-1$) and LG2 (OAM$=+1$ and SAM$=-1$) beams are co-propagating and they interact with 1st and 2nd components of BEC, respectively.  Where the  ground states of 1st component and 2nd component of the  $^{87}$Rb BEC are $|  5s_{\frac{1}{2}} F=1, m_f =-1 \rangle$ and $| 5s_{\frac{1}{2}} F=2, m_f =+1 \rangle$, respectively. $\Delta=-1.5$ GHz represents two-photon detuning. T-1, T-2, T-3, t-1, t-2, and t-3 are two-photon transitions channels.}
\end{figure}

\begin{figure*}[!h]

\subfloat[]{\includegraphics[trim = 1cm 0.5cm 0.1cm 1.5cm, scale=.30]{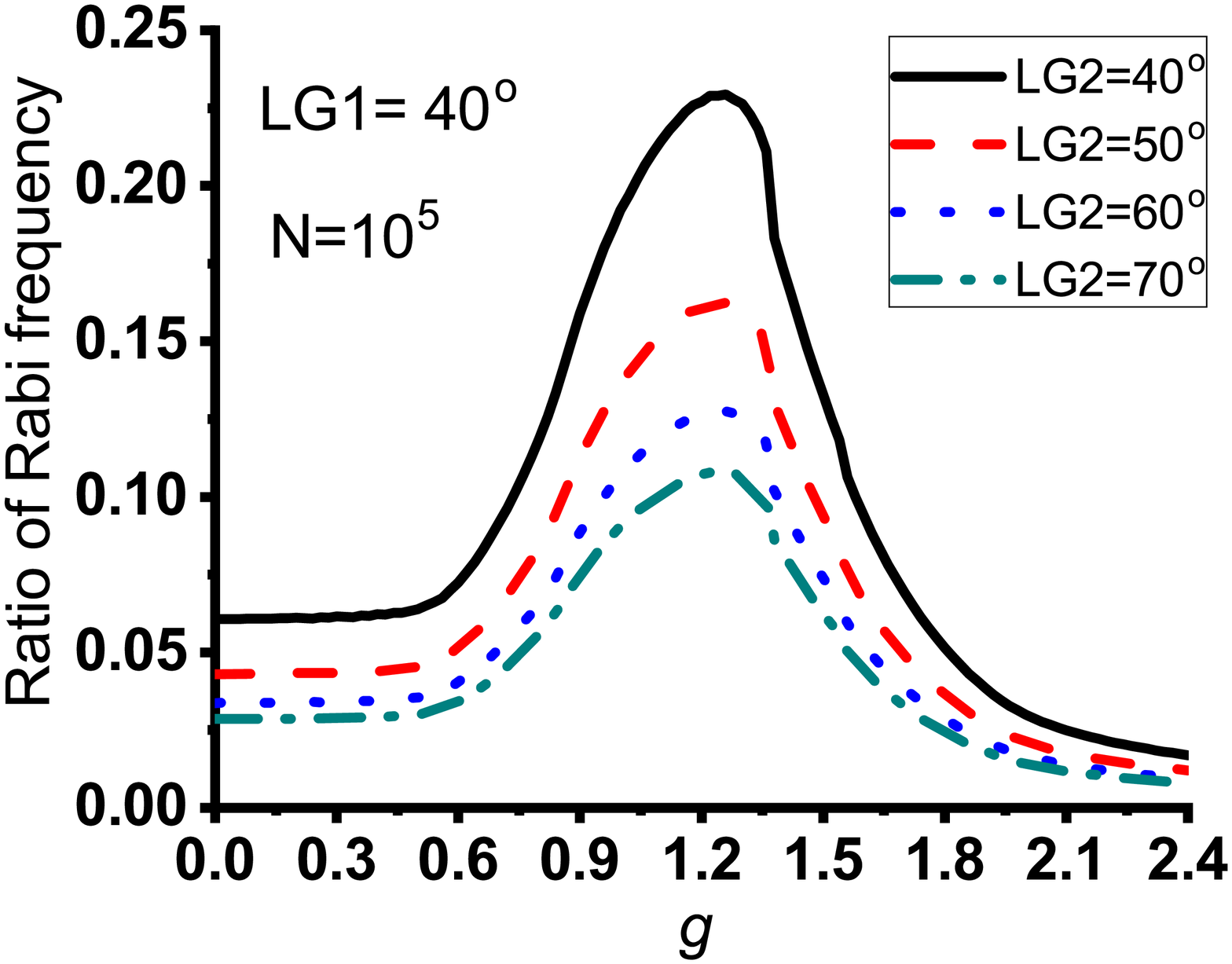}}
\subfloat[]{\includegraphics[trim = 1cm 0.5cm 0.1cm 1.5cm, scale=.30]{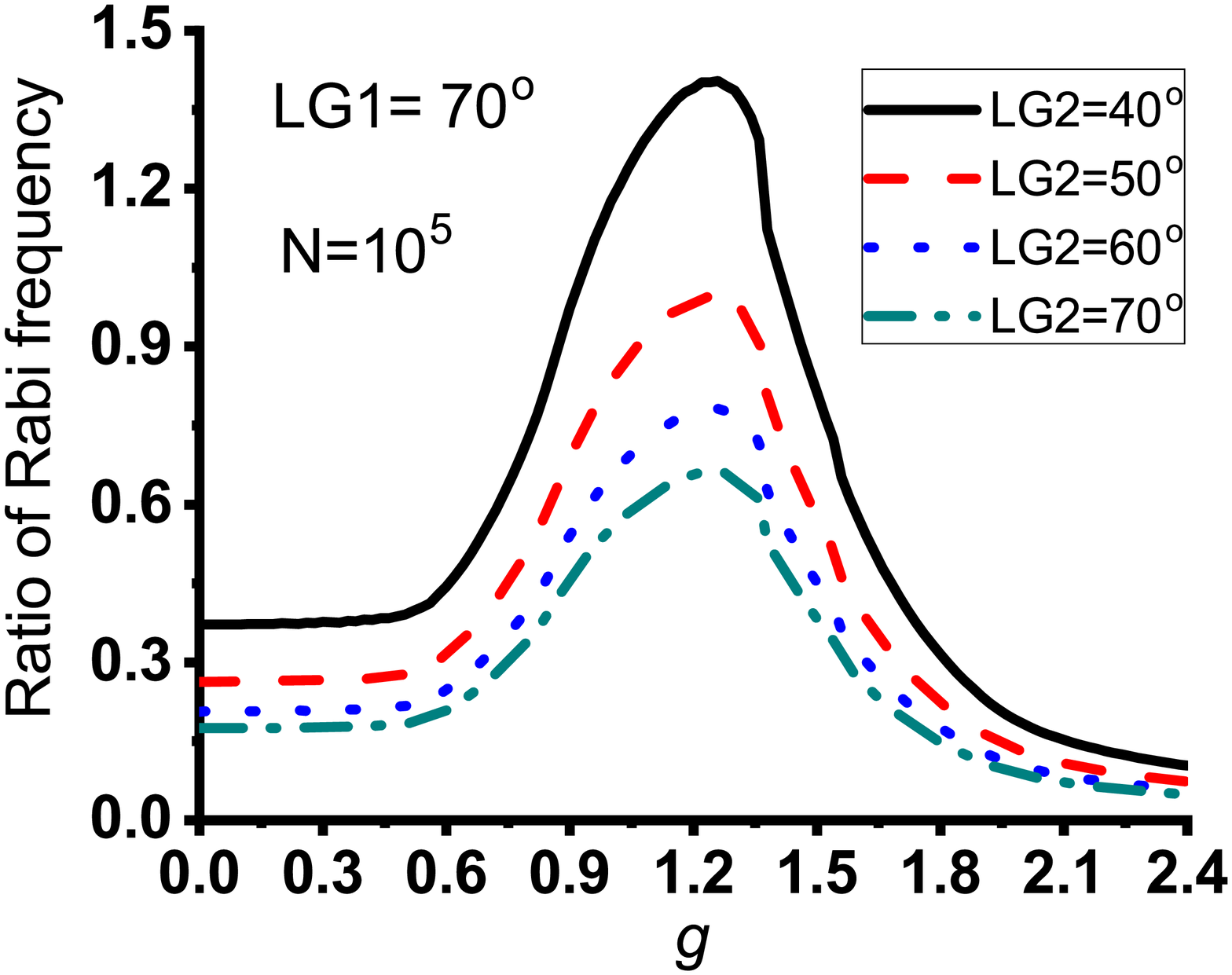}}\\
\subfloat[]{\includegraphics[trim = 1cm 0.5cm 0.1cm 1.5cm, scale=.30]{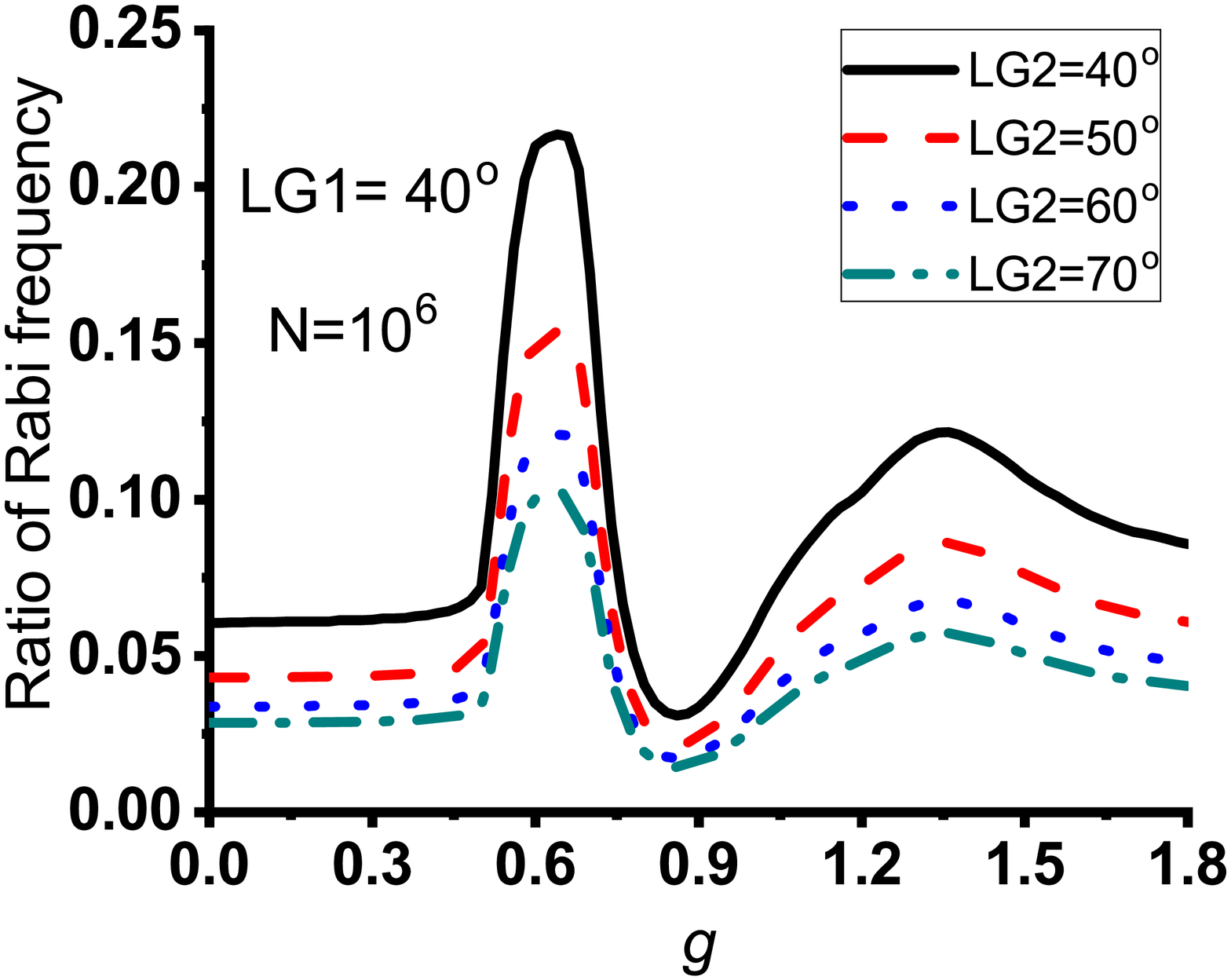}}
\subfloat[]{\includegraphics[trim = 1cm 0.5cm 0.1cm 1.5cm, scale=.30]{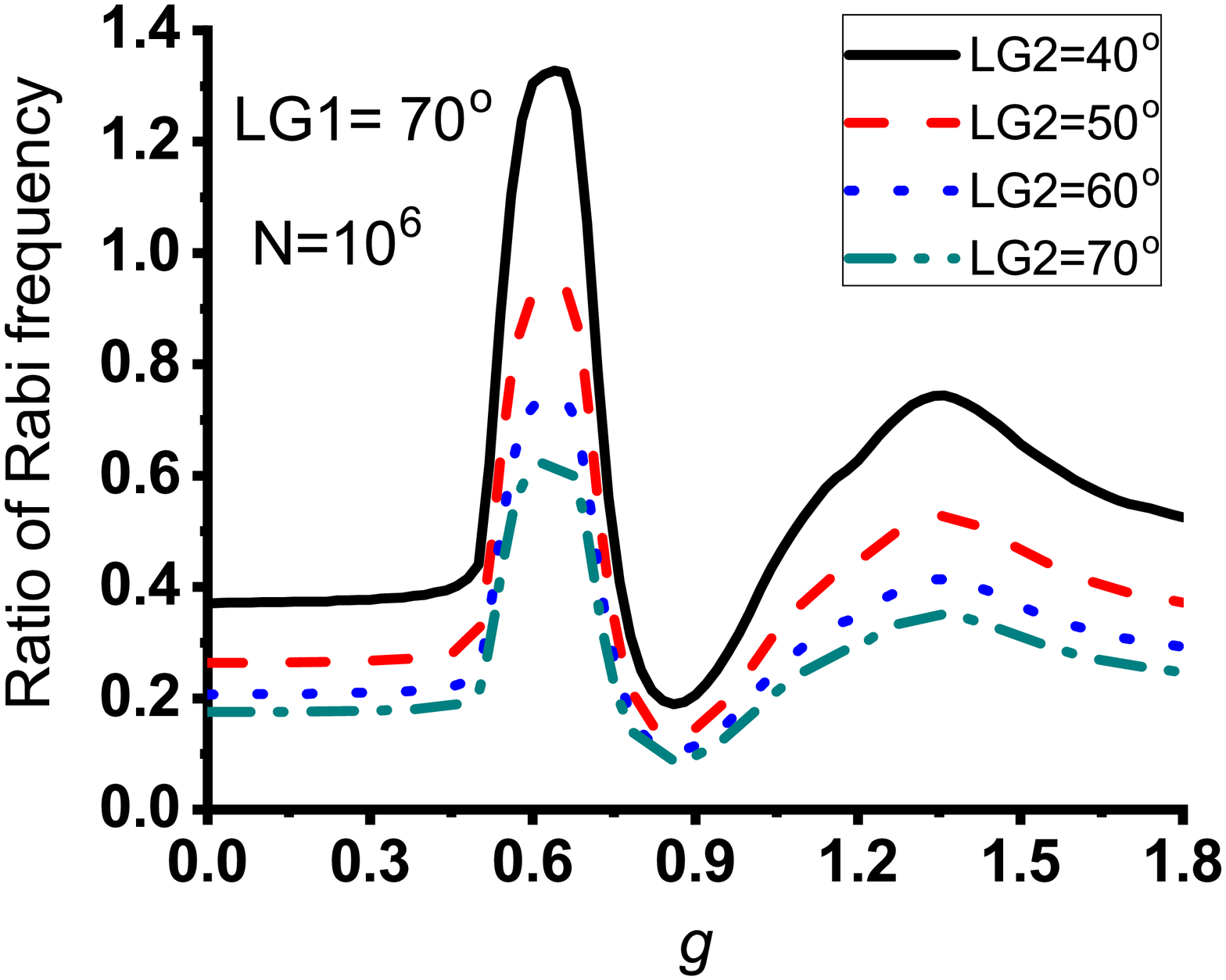}}
\caption{Variation of  ratio of Rabi frequencies of two-photon transition through T-2 and t-1  channels are plotted with respect to  inter-component interaction strength $(g)$  in case of $N=10^5$ and $10^6$. }
\end{figure*}

\subsection{Estimation of non-paraxial effect of LG beam through the  Rabi Frequencies}

To estimate the effects of the non-paraxial  signature of the LG beam in interaction with the two-component BEC, one of the best approaches is to compare the strength of the side-band transitions (say, T-2 )  with the prime transition (say, t-1). Let us now consider that  LG1 and LG2 are generated by focusing the paraxial beams with (OAM, SAM) = $(+1,-1)$ and $(+1,-1)$, respectively. FIG. 5 represents the schematic diagrams of the transitions among  the energy levels. Unlike the last case (see FIG. 1),  T-2 transition channel couples here with the t-1 transition channel via  $|  5p_{\frac{3}{2}} F'=2, m_f =0 \rangle$ as intermediate state. Since T-2 transition channel is only possible (between T-2 and t-1) through spin-orbit coupling of the focused LG1 beam, it is expected that the Rabi frequency profile due to this channel is strongly affected  by the focusing angle of the beam. Also, we consider that Gaussian beam is used to take the atoms back to BEC-2 only when intermediate state is $|  5p_{\frac{3}{2}} F'=2, m_f =0 \rangle$. Therefore,  atoms, which are excited through other T-1, T-3, t-2 and t-3 channels, will be lost from the trap.  Since the above selection of OAM and SAM of LG1 beam transfer $-1$ unit of OAM to the atoms at BEC-1 through the T-2 channel,  a vortex-antivortex superposed state is created at the electronic state of BEC-2, i.e, $|5s_{\frac{1}{2}} F=2, m_f =+1 \rangle$, with the help of two-photon Raman transition. Interestingly, the density distribution of this vortex and anti-vortex depends on the initial non-vortex structure of the BEC-2 and BEC-1, respectively.

The effect of the non-paraxial nature of the LG beam on this interaction for different density of atoms can be explicit from the distribution of the  ratio of Rabi frequencies  of Raman transitions through T-2 and t-1 channels in FIG. 6. For example, we consider the two different populations of BEC with $N= 10^5$ and $10^6$. In the figures FIG. 6(a,c) and 6(b,d), we consider that the LG1 beam is focused at an angle 40$^\circ$ and  70$^\circ$, respectively. In the figures, the focusing angle of LG2 is varied from 40$^\circ$ to 70$^\circ$. The structures of the distributions definitely  depend on the inter-component coupling through the initial density distributions of the BEC components.  It is clear from the figures that the ratio attains a maximum value at around $g=0.64$ for $N=10^6$ and  $g=1.25$ for $N=10^5$ in unit of 5.5 nm. Therefore, it is possible to enhance side-band transition significantly over primary transitions even with a comparatively low focused LG beam, if we choose the inter-component coupling strength properly. This phenomenon has a large impact on any experiments where non-paraxial vortex beam is or can be used \cite{Zhang2018}.  It is obvious in the figure that at $g=0$ (when the components of BEC are independent of each other), strong focusing is the only possibility to increase the non-paraxial effects of the LG beam. Here, in the case of the coupled-BEC, the effects can also be controlled by specifying  the  intra- and inter- component interactions of the BECs. In fig 6(b,d), one can see that  T-2 (which arises due to the spin-orbit coupling of light) even attains some values which is large compared to the prime transition t-1, which can not be possible in case of one-component BEC. 

It is possible to carry out experimental study of the above effect of the non-paraxial nature in the interaction with the two-component BEC. In the above scheme of creating the vortex-antivortex superposition from the side-band and primary transitions at the energy level $|5s_{\frac{1}{2}} F=2, m_f =+1 \rangle$ generates an interference pattern \cite{Bhowmik2016}. Let us consider that LG1 and LG2 beams are focused at angles 70$^\circ$  and 40$^\circ$, respectively, for $N=10^6$ atoms. We choose these particular combination of focusing angles of LG beams as they significantly affect the side-band transitions (see FIG. 6). The interference pattern displayed in FIG. 7 for $g=0$ and $g=0.64$ is in the $Z=0$ plane. At $g=0$, the populations of anti-vortex state through T-2 transition channel is much smaller compared to the population of vortex state through t-1 transition channel. In case of $g=0.64$, the situation is opposite and both the populations from T-2 and t-1 transition are closer to each other as well. Therefore, near to maximally coherent interference pattern is produced as shown in 7(b). After fixing the focusing angles, if we tune the inter-component coupling strength of the BEC mixture, we will be able to estimate $g$-value when effect of the non-paraxial nature of the vortex beam is maximal.

\begin{figure*}[!h]

\subfloat[]{\includegraphics[trim = 2cm 1.0cm 0.1cm 1.5cm, scale=.15]{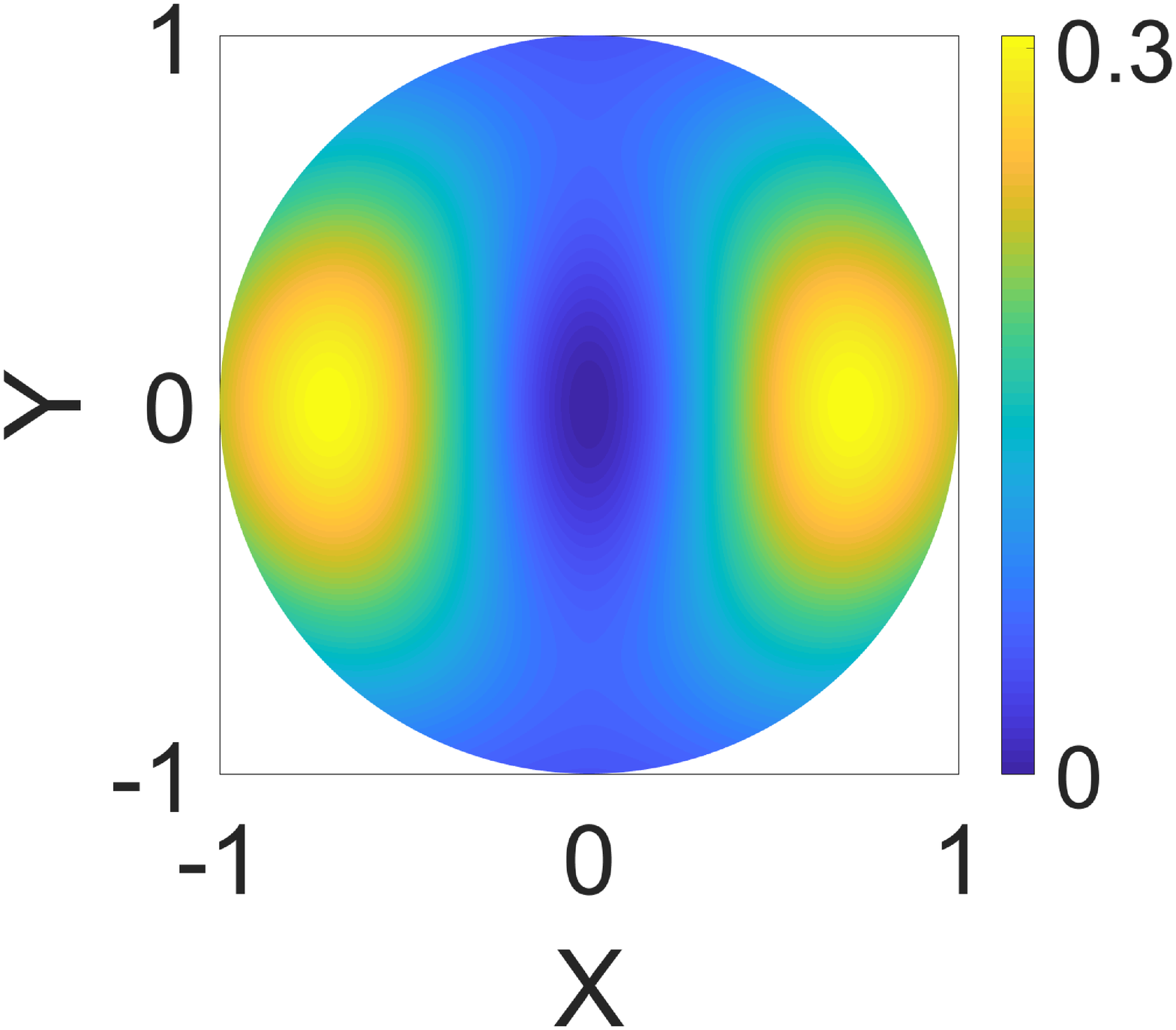}}
\subfloat[]{\includegraphics[trim = 1cm 1.0cm 0.1cm 5.5cm, scale=.15]{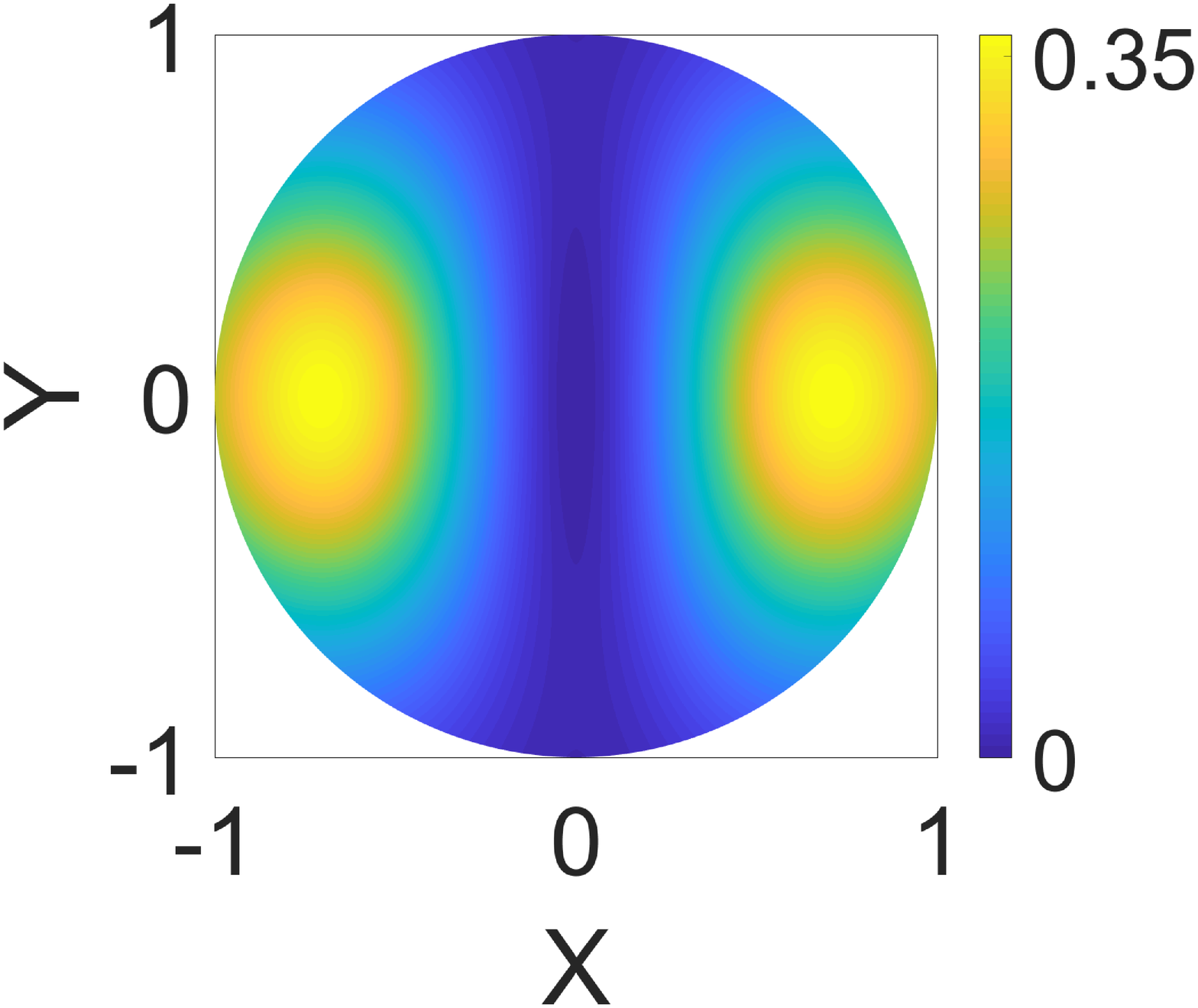}}
\caption{Images of superposition of vortex-antivortex for $N=10^6$ atoms at $Z=0$ plane are displayed. LG1 and LG2 are focused are considered at focusing angles  70$^\circ$  and 40$^\circ$, respectively.  Fig (a) is for $g=0$ and (b) is for $g=0.64$.  }
\end{figure*}

\section{CONCLUSION}

We have formulated a theory of interaction of the two-component BEC with the Laguerre-Gaussian (LG) beam, which is beyond the paraxial limit. Due to the coupling of the orbital and spin angular momentum of the light, the interaction of the focused LG beam with each of the components of the BEC takes place under three different angular momentum channels. Using the two photon Raman transitions, we calculate the Rabi frequencies of these angular momentum channels and show the variation of the Rabi frequency with the inter-BEC interaction strengths   and focusing angles of the  beam. We demonstrate the estimation procedure of the phase separation between the initial structure of the components of the BEC from the profiles of the Rabi frequencies. We have seen that the strengths of the side-band transitions achieve a maximum value for a particular value of the inter-BEC interaction strength and that is even larger than that of the strength of the primary transition for larger focusing angle of the  beam. An experimental scheme is
proposed to estimate the inter-coupling strength of the binary BEC by observing the coherence of the interference pattern based on the vortex-antivortex superposition from the side-band and primary transitions. Considering different orbital angular momentum of the incident beam and using the angular momentum channels, one can create multiply quantized vortices in the two-component BEC \cite{Kuopanportti2012, Kuopanportti2015}. These novel phenomenon of occurring  multiply quantized vortices in BEC is observed in multicomponent superconductivity \cite{Milosevic2015} or in rotating two-band Fermi gas \cite{Klimin2018}. The vortex-antivortex superpositions appear as the counter-rotating persistent currents in superconducting
circuits \cite{Nakamura1999, Friedman2000} which are propitious candidates for qubits in quantum-information processing and quantum communication networks \cite{Spedalieri2006}. Also, the vortices and multiple-vortices have been the subject of intensive experimental research in trapped superfluid Fermi gases \cite{Zwierlein2005, Zwierlein2006a, Zwierlein2006b} and even in real  condensed matter system \cite{Chmiel2018}.  We believe this is one of the best approaches to study the effect of the non-paraxial nature of the vortex beam on ultra-cold atoms. This non-paraxial effects of the angular momentum channels could be experimentally verified by measuring the orbital angular momentum in the components of BEC using surface wave spectroscopy \cite{Chevy2000, Haljan2001}.

\section*{ACKNOWLEDGMENTS}

We thank Rohit Kishan Ray, IIT Kharagpur for useful comments on the manuscript.
\clearpage




\end{document}